\documentclass[a4paper,11pt]{article}
\pdfoutput=1 

\usepackage[T1]{fontenc} 
\usepackage{amsmath,amssymb,mathtools}
\usepackage{color}
\usepackage{cite}
\usepackage[colorlinks=true,linkcolor=blue, citecolor=red, urlcolor=blue, bookmarks]{hyperref}
\usepackage{graphics}
\definecolor{comments}{RGB}{220,20,60}
\usepackage[text={17.1cm,24.6cm},centering]{geometry} 
\usepackage{braket}
\usepackage{comment}
\usepackage{tikz}

\usepackage{authblk}
\usepackage{cleveref}
\usepackage{multirow,makecell}
\usepackage{textcomp}
\usepackage{wasysym}
\numberwithin{equation}{section}

\newcommand{\beq}{\begin{equation}}
\newcommand{\eeq}{\end{equation}}
\newcommand{\bea}{\begin{eqnarray}}
\newcommand{\eea}{\end{eqnarray}}

\newcommand{\nc}{\newcommand}
\nc{\ir}{\mathrm{i}}
\nc{\dd}{\mathrm{d}} 
\nc{\eE}{\mathrm{e}}
\nc{\Tr}{\text{Tr}}
\nc{\id}{\mathbb{I}}
\nc{\tdet}{\tilde{\det}}
\nc{\J}{\mathcal{J}}

\begin{document} 

\title{\bf Dynamics of charge-imbalance-resolved entanglement negativity after a quench in a free-fermion model}

\author[1]{Gilles Parez}
\affil[1]{\it Centre de Recherches Math\'ematiques (CRM),
Universit\'e de Montr\'eal,
P.O. Box 6128, Centre-ville Station,
Montr\'eal (Qu\'ebec), H3C 3J7,
Canada 
}

\author[2]{Riccarda Bonsignori}
\affil[2]{\it Ru\dj er Bo\v{s}kovi\'c Institute, Bijeni\v{c}ka cesta 54, 10000 Zagreb, Croatia}

\author[3,4]{Pasquale Calabrese}
\affil[3]{\it International School for Advanced Studies (SISSA) and INFN, Via Bonomea 265, 34136 Trieste, Italy}
\affil[4]{\it International Centre for Theoretical Physics (ICTP), Strada Costiera 11, 34151 Trieste, Italy}

\date{}
\maketitle

\begin{abstract}
The presence of a global internal symmetry in a quantum many-body system is reflected in the fact that the entanglement between its subparts is endowed with an internal structure, namely it can be decomposed as sum of contributions associated to each symmetry sector. 
The symmetry resolution of entanglement measures provides a formidable tool to probe the out-of-equilibrium dynamics of quantum systems. 
Here, we study the time evolution of charge-imbalance-resolved negativity after a global quench in the context of free-fermion systems, complementing former 
works  for the symmetry-resolved entanglement entropy.

We find that the charge-imbalance-resolved logarithmic negativity shows an effective equipartition in the scaling limit of large times and system size, with a perfect equipartition for early and infinite times. We also derive and conjecture a formula for the dynamics of the charged R\'enyi logarithmic negativities. We argue that our results can be understood in the framework of the quasiparticle picture for the entanglement dynamics, and provide a conjecture that we expect to be valid for generic integrable models. 

\end{abstract}

\baselineskip 18pt
\thispagestyle{empty}
\newpage

\tableofcontents

\section{Introduction}

The nonequilibrium dynamics of isolated quantum systems received considerable attention over the last two decades. In particular, the entanglement dynamics plays a fundamental role in our understanding of numerous aspects of quantum many-body systems out of equilibrium, such as the equilibration and thermalisation of isolated many-body systems \cite{PolkonikovRMP11, GE15, DKPR15, SI,EF16}, the emergence of thermodynamics in quantum systems \cite{C:18, DLS:13, SPR:11,ckc-14} or the effectiveness of classical computers to simulate quantum dynamics \cite{SWVC:PRL, SWVC:NJP, PV:08, HCTDL:12, D:17}. In one-dimensional quantum integrable systems, the entanglement dynamics after a  quantum quench \cite{cc-05,cc-06,cc-07}, the simplest and most broadly studied protocol to drive a quantum system out of equilibrium, is well described and understood in terms of the \textit{quasiparticle~picture}~\cite{cc-05, ac-17,ac-18,c-20}. These theoretical developments benefited from pioneering cold-atom and ion-trap experiments that could probe isolated quantum systems at large time scales with an unprecedented precision. Most notably, it has been possible to measure the entanglement of many-body systems out of equilibrium \cite{kaufman-2016,exp-lukin,brydges-2018,ekh-20}. 
%
%

For a bipartite quantum system in a pure state described by a density matrix $\rho$, the entanglement between a subsystem $A$ and its complement $B$ is quantified by the \textit{R\'enyi entropies}
\begin{equation}
    \label{RenyiEnt}
S_n \equiv\frac{1}{1-n}\log \Tr \rho_A^n, 
\end{equation}
where $\rho_A = \Tr_{B}\rho$ is the reduced density matrix (RDM) of the subsystem $A$. In particular, the limit $n\to 1$ of the R\'enyi entropies yields the celebrated \textit{entanglement entropy}
\begin{equation}
S_1\equiv \lim_{n\to 1}S_n = -\Tr \rho_A \log \rho_A.
\end{equation}

The R\'enyi entropies quantify the entanglement between a subsystem $A$ and its complement, irrespective of the topology of $A$. It is often relevant to consider the case where $A$ is itself a bipartite system $A=A_1 \cup A_2$, and investigate the entanglement between $A_1$ and $A_2$. To this end, one usually introduces the  \textit{R\'enyi mutual informations}, defined as
\begin{equation}
    \label{RenyiMutInfo}
    I_{n}^{A_1:A_2}= S^{A_1}_{n} +  S^{A_2}_{n} -S^{A_1\cup A_2}_{n}=\frac{1}{n-1}\log  \left(\frac{\Tr\rho_{A_1\cup A_2}^n}{\Tr\rho_{A_1}^n\Tr\rho_{A_2}^n} \right).
\end{equation}
However, these quantities are not proper entanglement measures but rather quantify the global correlations between the two subsystems \cite{wvhc-08}. 

A suitable measure of entanglement between two non-complementary subsystems $A_1$ and $A_2$ is instead the  \textit{entanglement negativity}, defined as  \cite{vw-02}
\begin{equation}
\label{eq:NegDef}
\mathcal{N}^{(b)}\equiv \frac{\Tr |\rho_A^{T_1}|-1}{2},
\end{equation}
where $\Tr |O|= \Tr \sqrt{O^{\dagger}O}$ is the trace norm of the operator $O$, and $\rho_A^{T_1}$ is the partially-transposed RDM of the system $A=A_1 \cup A_2$. 
The latter is defined as follow. We denote the Hilbert spaces corresponding to each subsystem by $\mathcal{H}_1$ and $\mathcal{H}_2$, with respective bases $|e_j^1 \rangle$ and $| e_k^2\rangle$. With this notation, $\rho_A$ reads 
\begin{equation}
\label{rdmBosonic}
\rho_A=\sum_{ijkl} \langle e_i^1, e_j^2 | \rho_A |e_k^1, e_l^2 \rangle |e_i^1, e_j^2 \rangle \langle e_k^1, e_l^2|,
\end{equation}
and its partial transpose with respect to the degrees of freedom of subsystem $A_1$ is defined as
\begin{equation}
\label{PartialTranspose}
\rho_A^{T_1} \equiv \sum_{ijkl} \langle e_k^1, e_j^2 | \rho_A |e_i^1, e_l^2 \rangle |e_i^1, e_j^2 \rangle \langle e_k^1, e_l^2|.
\end{equation}
In terms of projectors on basis states, the partial transposition corresponds to the operation 
\begin{equation}
\left( |e_i^1,e_j^2 \rangle \langle e_k^1, e_l^2 | \right)^{T_1} \equiv| e_k^1, e_j^2 \rangle \langle e_i^1, e_l^2 |.
\end{equation}

The negativity is related to the existence of negative eigenvalues in the spectrum of the partially-transposed RDM. Indeed, writing $\Tr |\rho_A^{T_1}|$ in terms of the eigenvalues $\lambda_i$ of $\rho_A^{T_1}$, we have
\begin{equation}
    \mathrm{Tr}|\rho_A^{T_1}|=\sum_i |\lambda_i|=\sum_{\lambda_i >0}|\lambda_i|+\sum_{\lambda_i <0}|\lambda_i|=1+2\sum_{\lambda_i<0}|\lambda_i|,
\end{equation}
and hence 
\begin{equation}
\mathcal{N}^{(b)} = \sum_{\lambda_i<0}|\lambda_i|,
\end{equation}
in agreement with Peres separability criterion \cite{peres-1996,s-00,plenio-2005}.
We also introduce the  \textit{logarithmic negativity} as
\begin{equation}
\label{eq:Eb}
\mathcal{E}^{(b)}=\log \Tr |\rho_A^{T_1}|.
\end{equation}
When the system $A_1 \cup A_2$ is in a pure state, it satisfies $\mathcal{E}^{(b)}=S_{1/2}^{A_1}$ \cite{vw-02}.
However, the experimental measure of the negativity in many-body systems is nowadays only possible by full quantum tomography. 
For this reason, a few protocols to measure the moments $\mathrm{Tr}(\rho_A^{T_1})^n$ of the partial transpose have been proposed \cite{gbbb-18,csg-19,ekh-20,ncv-21},
leading to an actual  measure in an ion-trap experiment \cite{ekh-20,ncv-21}.
Some linear combinations of these moments provide sufficient conditions (known as $p_n$-PPT conditions) to witness entanglement in mixed states.
Hence, it is very important to provide theoretical predictions not only for the negativity, but also for the moments of the partial transpose 
(also known as R\'enyi negativities, see below). 

A very recent line of research concerns the understanding of the interplay between symmetries and entanglement out of equilibrium, as highlighted in a recent experiment \cite{exp-lukin}. 
In the case where the system has a global symmetry, the entanglement splits between the various symmetry sectors, and this \textit{symmetry resolution of entanglement} 
attracted a lot of attention recently \cite{lr-14,GS,equi-sierra,cgs-18,bons,fg-19,Anyons,bons-20,fg-20,Luca,mdc-20,mdc-20b,ccdm-20,tr-20,mrc-20,trac-20,Topology,hc-20,as-20,pbc-21,pbc-21bis,mbc-21,hcc-21,vecd-21,fg-21,
ncv-21,eimd-21,c-21,cc-21,hcc-21b,chcc-21,cdm-21,znm-21,wznm-21,ms-21,c-21b,ore-21,fkeg-22,AMC22,J22,G22, CCADFMSS22, PVCC22, BBCG22}. Moreover, symmetry-resolved quantities have already been measured experimentally \cite{vecd-21,ncv-21}. 
In the context of non-complementary subsystems and mixed-states, the \textit{charge-imbalance-resolved negativity} was defined in \cite{cgs-18} and has since been investigated 
in several circumstances \cite{mbc-21,ms-21,c-21b,ncv-21}. 
It plays a crucial role in the detection of entanglement in mixes-state quantum many-body systems \cite{ncv-21}. 
However, little results are available regarding its non-equilibrium dynamics \cite{ncv-21}. 
In this paper, we fill this gap studying  two quenches from homogeneous initial states in a free-fermion model.
In particular, we interpret our results in terms of the quasiparticle picture for the entanglement dynamics, and generalise results for the R\'enyi negativities \cite{mac-21}. 

This paper is organised as follows. In Sec. \ref{sec_cir_def} we introduce a definition of negativity for fermionic systems based on the  partial time-reversal (TR) transformation of the RDM. We also discuss the charge-imbalance-resolved negativity and its expression in terms of Fourier transforms of the charged moments. We express these charged moments in terms of the two-point correlation matrix in the context of free fermions in Sec. \ref{sec:FF}. We also define the tight-biding model and the two initial states that we consider in this work. In Sec. \ref{sec:Nna}, we present our analytical results and conjectures for the quench dynamics of the charged moments for both quenches under consideration. We compute the Fourier transforms of the charged moments in Sec. \ref{sec:cir_t} and investigate the dynamics of the charge-imbalance-resolved negativity. Moreover, we argue that we recover known results for the dynamics of the total negativity from the charge-imbalance-resolved one. We interpret our results in terms of the quasiparticle picture for the entanglement dynamics in Sec. \ref{sec:QPP}, and present our conclusions and outlooks in Sec. \ref{sec:ccl}.

\section{Negativity for fermionic systems and charge imbalance}
\label{sec_cir_def}

In this section, we define the (fermionic) negativity and its charge-imbalance-resolved version in the context of fermionic systems with a global $U(1)$ symmetry. 

\subsection{Fermionic negativity}

The definition of the negativity in Eq. \eqref{eq:NegDef} is not well-suited to investigate entanglement properties in the context of free-fermion systems. The main reason is that, for such systems, the partial transpose $\rho_A^{T_1}$, unlike $\rho_A$, is not a Gaussian operator, but rather a sum of two non-commuting Gaussian operators~\cite{ez-15}. 
As a consequence, its full spectrum is not accessible \cite{ctc-15,ctc-15b,eez-18}. 
To circumvent this issue, the partial TR transformation of the RDM, denoted $\rho_A^{R_1}$, has been introduced \cite{ssr-17,ssr1-17,ssgr-18,sr-19,ryu,paola,ksr-20,smr-21,mvdc-22}.
To define the partial TR transformation, let us consider the example of a single-state system described by fermionic operators $c$ and $c^{\dagger}$, with $\{c, c^{\dagger}\}=1$. 
In the basis of the fermionic coherent states $|\xi\rangle = \eE^{-\xi c^{\dagger}}|0\rangle$ and $\langle \bar{\xi}|=\langle 0 | \eE^{-c \hspace{.05cm} \bar{\xi}}$, 
where $\xi, \bar{\xi}$ are Grassmann variables, the TR transformation is defined as 
\begin{equation}\label{eq:TR}
(\ket{\xi}\bra{\bar{\xi}})^R \equiv\ket{\ir\bar{\xi}}\bra{\ir\xi}.
\end{equation}
In particular, the TR transformation differs from the standard transposition by the presence of the factor~$\ir$. 
In the case of a many-particle system, the partial TR transformation on the degrees of freedom of subsystem $A_1$ is defined as
\begin{equation}\label{eq:mb}
    (\ket{ \{ \xi_j \}_{j\in A_1} ,\{ \xi_j \}_{j\in A_2}}\bra{\{ \bar{\chi}_j \}_{j\in A_1} ,\{ \bar{\chi}_j \}_{j\in A_2}})^{R_1}  \equiv \ket{ \{ \ir\bar{\chi}_j \}_{j\in A_1}, \{ \xi_j \}_{j\in A_2}}\bra{\{ \ir\xi_j \}_{j\in A_1}, \{ \bar{\chi}_j \}_{j\in A_2}},
\end{equation}
where $\ket{\{\xi_j\}}=\eE^{-\sum_j\xi_jc^{\dagger}_j}\ket{0}$, $\bra{\{\bar{\chi}_j\}}=\bra{0}\eE^{-\sum_j c_j\bar{\chi}_j}$ are the many-particle fermionic coherent states.

 From $\rho_A^{R_1}$, one defines the  \textit{fermionic negativity} as
\begin{equation}
\mathcal{N}\equiv  \frac{\mathrm{Tr}|\rho_A^{R_1}|-1}{2}= \frac{\mathrm{Tr} \sqrt{\rho_A^{R_1}(\rho_A^{R_1})^{\dagger}}-1}{2},
\label{NF}
\end{equation}
and the \textit{fermionic logarithmic negativity} is  
\begin{equation}
\label{eq:def_log_neg}
\mathcal{E}=\log \Tr |\rho_A^{R_1}|.
\end{equation}
It is possible to show that these fermionic negativities are entanglement monotones \cite{sr-19} and that they can capture some entanglement that is overlooked 
by the standard negativity. 
Since we only consider fermionic systems, we refer to these quantities as the negativity and the logarithmic negativity, respectively. In contrast, the quantities $\mathcal{N}^{(b)}$ and $\mathcal{E}^{(b)}$ in Eqs. \eqref{eq:NegDef} and \eqref{eq:Eb} are sometimes referred to as bosonic negativity and bosonic logarithmic negativity, respectively. Contrarily to the bosonic negativity, the (fermionic) negativity is not related to the presence of negative eigenvalues in the spectrum of the partial TR transformed RDM, see \cite{paola}.

We also introduce the R\'enyi (logarithmic) negativities as \cite{paola}
\begin{equation}\label{eq:RN}
\mathcal{N}_n=\frac{N_{n}-1}{2},\qquad 
\mathcal{E}_n = \log N_n, \qquad
N_n= 
\begin{cases}
  \mathrm{Tr}\big(\rho_A^{R_1}(\rho_A^{R_1})^{\dagger}\dots  \rho_A^{R_1}(\rho_A^{R_1})^{\dagger}\big), &\quad \mathrm{even} \ n,\\[.2cm]
 \mathrm{Tr}\big(\rho_A^{R_1}(\rho_A^{R_1})^{\dagger}\dots  \rho_A^{R_1}\big),&  \quad \mathrm{odd} \ n,
\end{cases}
\end{equation}
from which the negativity and logarithmic negativity are obtained in the replica limit $n_e\to 1$ with even~$n_e$ \cite{cct-12,cct-13},
\begin{equation}
 \mathcal{N}= \lim_{n_e \to 1}  \mathcal{N}_{n_e}, \quad \mathcal{E} = \lim_{n_e \to 1}  \mathcal{E}_{n_e}.
\end{equation}
We stress that the limit $\lim_{n_e \to 1}  N_{n_e}=\mathrm{Tr} \sqrt{\rho_A^{R_1}(\rho_A^{R_1})^{\dagger}}$ is different from 
$N_1 = \Tr (\rho_A^{R_1})=1$.

\subsection{Charge-imbalance resolution}
 
We consider an extended quantum system with an internal $U(1)$ symmetry generated by a local charge $Q$. If the system is described by a density matrix $\rho$ that only acts non-trivially in an eigenspace of $Q$, we have $[\rho,Q]=0$. In the case of a bipartition in two complementary subsystems $A$ and $B$, by locality, the charge $Q$  splits as sum of operators that act on the local degrees of freedom of the two parts, $Q=Q_A+Q_B$. Hence, the trace over the degrees of freedom of subsystem $B$ in the commutation relation $[\rho,Q]=0$ yields $[\rho_A,Q_A]=0$, i.e. $\rho_A$ has a block diagonal form with each block corresponding to an eigenvalue of $Q_A$.

In the case where the subsystem $A$ is itself partitioned into two complementary subsystems $A_1$ and $A_2$, we denote with $Q_1$ and $Q_2$ the corresponding charge operators. As shown in \cite{cgs-18}, the partial transpose with respect to the degrees of freedom of subsystem $A_1$ performed on the commutation relation $[\rho_A, Q_A]=0$ gives $[\rho_A^{T_1},Q_2-Q_1^{T_1}]=0$,
so that the partially-transposed RDM can be decomposed in blocks that correspond to the eigenvalues of the charge-imbalance operator $Q_2-Q_1^{T_1}$.
As pointed out in \cite{mbc-21}, a similar relation holds for the partial TR transformation, namely
\begin{equation}
\label{eq:RDM-dec}
[\rho_A^{R_1},Q_2-Q_1^{R_1}]=0.
\end{equation}
In the following, we define the operator $\hat{Q}_A\equiv Q_2-Q_1^{R_1}$ as the charge-imbalance operator, since we only deal with fermionic systems and partial TR transformation. We denote the eigenvalues of $\hat{Q}_A$ by $q$, and the projector on the corresponding eigenspace is $\Pi_q$. 
The charge-imbalance operator is basis-dependent (see, e.g., \cite{mbc-21}), and reads $\hat{Q}_A=Q_2+Q_1-\ell/2$ in our computational basis, where $\ell$ is the length of the subsystem $A$, see \cite{mbc-21}. 
From Eq.~\eqref{eq:RDM-dec}, we have the  following decomposition of the partial TR transformed RDM,
\begin{equation}
    \rho_A^{R_1}=\oplus_{q}p(q)\rho_A^{R_1}(q),
\end{equation}
where  $p(q)=\mathrm{Tr}(\Pi_{q}\rho_A^{R_1})$ is the probability of finding $q$ as the outcome of a measurement of $\hat{Q}_A$. The operator $\rho_A^{R_1}(q)$ is the \textit{charge-imbalance-resolved partial TR transformed RDM}, defined as
\begin{equation}
    \rho_A^{R_1}(q)=\frac{\Pi_{q}\rho_A^{R_1}\Pi_{q}}{\mathrm{Tr}(\Pi_{q} \rho_A^{R_1})}, \qquad \mathrm{Tr}( \rho_A^{R_1}(q))=1.
\end{equation}
%
The \textit{charge-imbalance-resolved  negativity} is defined as
\begin{equation}
    \mathcal{N}(q)=\frac{\mathrm{Tr}|\rho_A^{R_1}(q)|-1}{2}, \qquad 
\end{equation}
and satisfies
\begin{equation}
\label{eq:Negtot}
\mathcal{N}=\sum_{q }p(q)\mathcal{N}(q).
\end{equation} 
Here we also define a \textit{charge-imbalance-resolved logarithmic negativity} as
\begin{equation}
\hat{\mathcal{E}}(q) = \log \mathrm{Tr}|\rho_A^{R_1}(q)|= \log (2 \mathcal{N}(q)+1)\,.
\end{equation} 
However, because of the non-linearity of the logarithm, $\hat{\mathcal{E}}(q)$ does not represent a proper resolution (in the sense that it is not the contribution of the sector $q$ 
to the total logarithmic negativity), but it is a useful auxiliary quantity related to ${\mathcal{N}}(q)$.

It is also useful to define the {\it charge-imbalance-resolved R\'enyi negativities}
\begin{equation}\label{eq:CIR-RN}
\mathcal{N}_n(q) = \frac{\hat{N}_n(q)-1}{2}, \qquad 
\hat{N}_n(q)= 
\begin{cases}
  \mathrm{Tr}\big(\rho_A^{R_1}(q)(\rho_A^{R_1}(q))^{\dagger}\dots  \rho_A^{R_1}(q)(\rho_A^{R_1}(q))^{\dagger}\big), & \quad \mathrm{even} \ n,\\[.2cm]
 \mathrm{Tr}\big(\rho_A^{R_1}(q)(\rho_A^{R_1}(q))^{\dagger}\dots  \rho_A^{R_1}(q)\big),&  \quad \mathrm{odd} \ n,
\end{cases}
\end{equation} 
and we have $\displaystyle \mathcal{N}(q)= \lim_{n_e \to 1} \mathcal{N}_{n_e}(q)$.

To compute these quantities, the method developped in \cite{cgs-18} is similar to the case of the symmetry-resolved entropies \cite{GS}, where one introduces the charged moments \cite{cmr-13,bhm-15,cnn-16} and investigates their Fourier transforms. To proceed, we introduce the \textit{charged R\'enyi logarithmic negativities} \cite{mbc-21}
\begin{equation}\label{eq:supalpha}
\mathcal{E}_n(\alpha) = \log N_n(\alpha), \qquad N_n(\alpha)= 
\begin{cases}
  \mathrm{Tr}(\rho_A^{R_1}(\rho_A^{R_1})^{\dagger}\dots  \rho_A^{R_1}(\rho_A^{R_1})^{\dagger}\eE^{\ir \hat{Q}_A\alpha}), & \quad \mathrm{even} \ n,\\
 \mathrm{Tr}(\rho_A^{R_1}(\rho_A^{R_1})^{\dagger}\dots \rho_A^{R_1}\eE^{\ir \hat{Q}_A\alpha}), & \quad \mathrm{odd} \ n,
\end{cases}
\end{equation}
where the quantities $N_n(\alpha)$ are the charged moments. The limit $n_e \to 1$ defines the \textit{charged logarithmic negativity} 
\begin{equation}
\label{eq:def_Ea}
\mathcal{E}(\alpha) \equiv \lim_{n_e \to 1} \log N_{n_e}(\alpha),
\end{equation}
and we call the moment $N_1(\alpha)=\Tr [\rho_A^{R_1}\eE^{\ir \hat{Q}_A\alpha}]$ the \textit{charged probability}, since its Fourier transform gives the probability $p(q)$.
 
The Fourier transforms of the charged moments yield the charge-imbalance-resolved negativities \cite{mbc-21},
\begin{equation}
\label{eq:pqft}
    \mathcal{Z}_{R_1,n}(q)=\displaystyle \int_{-\pi}^{\pi}\frac{d\alpha}{2\pi }\eE^{- \ir q \alpha}N_{n}(\alpha), \quad p(q)=\displaystyle \int_{-\pi}^{\pi}\frac{d\alpha}{2\pi }\eE^{- \ir q \alpha}N_{1}(\alpha),
\end{equation}
from which 
\begin{equation}
\label{eq:rnq}
    \hat{N}_{n}(q)=\frac{\mathcal{Z}_{R_1,n}(q)}{p(q)^n}.
\end{equation}
We also introduce the quantity
\begin{equation}
\label{eq:ZR1}
\mathcal{Z}_{R_1}(q)\equiv\lim_{n_e \to 1} \mathcal{Z}_{R_1,n_e}(q) = \int_{-\pi}^{\pi}\frac{d \alpha}{2 \pi} \eE^{-\ir  q \alpha} \eE^{\mathcal{E}(\alpha)}, 
\end{equation}
and express the charge-imbalance-resolved (logarithmic) negativity as
\begin{equation}
\label{eq:NegDefNeel}
 \mathcal{N}( q) =\frac 12 \Big( \frac{\mathcal{Z}_{R_1}(q)}{ p( q)}-1\Big), \qquad \hat{\mathcal{E}}(q) = \log\Big( \frac{\mathcal{Z}_{R_1}(q)}{ p( q)}\Big).
\end{equation}

\section{Charged moments for free fermions}\label{sec:FF}

In this section, we introduce our quench protocol, i.e. the two initial states we consider and 
the free-fermion model that governs the time evolution. We also provide exact formulas for the charged moments in terms of the two-point correlation matrix.

\subsection{Model and initial states}

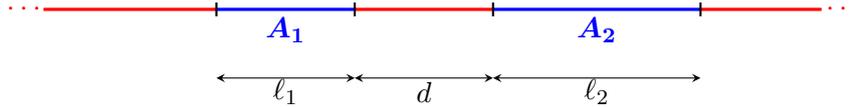
\begin{figure}
\begin{center}
  \begin{tikzpicture}[>=stealth, scale=0.91]

      \draw [-,red,very thick] (-0.5,0)--(2,0);
      \draw[red,thick] (-0.75, 0) node{$\cdots$};    
   \draw [-,blue,very thick] (2,0)--(4,0);   
     \draw [-,red,very thick] (4,0)--(6,0);   
       \draw [-,blue,very thick] (6,0)--(9,0);
      \draw [-,red,very thick] (9,0)--(10.75,0);
      \draw[red,thick] (11.1, 0) node{$\cdots$};

           \draw[thick](2,-0.1)--(2,0.1);
           \draw[thick](4,-0.1)--(4,0.1);
            \draw[thick](6,-0.1)--(6,0.1);
             \draw[thick](9,-0.1)--(9,0.1);
           
            \draw[blue, thick] (3, -0.3) node{ $\boldsymbol{A_1}$};
              \draw[blue, thick] (7.5, -0.3) node{ $\boldsymbol{A_2}$};

           \draw[<->] (2,-1)--(4,-1);
            \draw (3, -1.2) node{$\ell_1$};
            
             \draw[<->] (4,-1)--(6,-1);
            \draw (5, -1.2) node{$d$};
            
             \draw[<->] (6,-1)--(9,-1);
            \draw (7.5, -1.2) node{$\ell_2$};

   \end{tikzpicture}  
\end{center}
\caption{Illustration of the geometry we consider in the limit where the system size $L$ is very large. The system $A=A_1\cup A_2$ is in blue, whereas the system $B$ consists of the red regions.}
\label{fig:Disjoint}
\end{figure}

In the following, we study the time evolution of the charged moments and the charge-imbalance-resolved negativity after a global quench in the tight-binding model with Hamiltonian
\begin{equation}
\label{eq:Hfree}
H=\sum_{j=1}^{L}(c_j^{\dagger}c_{j+1}+c_{j+1}^{\dagger}c_{j}). 
\end{equation}
Here, $c_j$ and $c_j^\dagger$ are the canonical fermionic annihilation and creation operators on site $j$. They satisfy the anticommutation relations $\{c_i,c_j^\dagger \} = \delta_{i,j}$ and $\{c_i,c_j \} = \{c_i^\dagger,c_j^{\dagger}\}=0$. The system size is $L$, and we consider a chain with periodic boundary conditions. For simplicity, we assume that $L$ is even. The conserved charge is the fermion-number operator
\begin{equation}
    \label{eq:Qff}
    Q=\sum \limits_{j=1}^L c_j^\dagger c_j.
\end{equation}

In the chain, we consider a bipartition $A \cup B$ where $A$ consists of two intervals $A_1$ and $A_2$ separated by $d$ lattice sites, of respective lengths $\ell_1$ and $\ell_2$, with $\ell_1+\ell_2=\ell$. We illustrate this geometry in Fig. \ref{fig:Disjoint}. For simplicity, we also assume that $d,\ell_1,\ell_2$ are even numbers. The global conserved charge $Q$ trivially splits as a sum over $A_1$, $A_2$ and $B$, 
\begin{equation}
Q = \sum_{j \in A_1}c_j^\dagger c_j +\sum_{j \in A_2}c_j^\dagger c_j+\sum_{j \in B}c_j^\dagger c_j  \equiv Q_1+Q_2+Q_B.
\end{equation}

We consider two simple homogeneous initial states, namely the N\'eel and the dimer states:
\begin{equation}
    \label{eq:InitialStates}
    \begin{split}
        |N\rangle&=  \prod_{j=1}^{L/2}c_{2j}^{\dagger}|0\rangle, \\
        |D\rangle &= \prod_{j=1}^{L/2}\frac{c_{2j}^{\dagger}-c_{2j-1}^{\dagger}}{\sqrt{2}}|0\rangle.
    \end{split}
\end{equation}
 These two states enjoy important properties. First, the time-dependent density matrix 
\begin{equation}
\rho(t) = \eE^{-\ir t H}|\psi_0\rangle \langle \psi_0| \eE^{\ir t H}, \qquad |\psi_0\rangle = |N\rangle,|D\rangle,
\end{equation}
where $H$ is the tight-biding Hamiltonian \eqref{eq:Hfree}, commutes with the total charge $Q$ in Eq. \eqref{eq:Qff} and is a Gaussian operator for all values of $t$. Second, their time-dependent correlation matrix 
\begin{equation}
[C(t)]_{x,x'}=\langle \psi_0| \eE^{\ir t H}c_x^\dagger c_{x'}\eE^{-\ir t H}|\psi_0\rangle,  \qquad |\psi_0\rangle = |N\rangle,|D\rangle,
\end{equation}
is exactly known. For the quench from the N\'eel state, we have \cite{ac-19}
\begin{equation}
\label{eq:CNeel}
[C(t)]_{x,x'}=\frac{\delta_{x,x'}}2 +\frac{(-1)^{x'}}2\int_{-\pi}^{\pi}\frac{\dd k}{2\pi}\eE^{\ir k(x-x')+4\ir t\cos k}.
\end{equation}
For the dimer state, it reads \cite{f-14}
\begin{subequations}
\label{eq:CDimer}
\begin{equation}
    [C(t)]_{j,k} = \Big\langle \begin{pmatrix}c_{2j-1}^\dagger \\ c_{2j}^\dagger \end{pmatrix} \begin{pmatrix}c_{2k-1} & c_{2k} \end{pmatrix} \Big\rangle = \frac 12 \big( \delta_{j,k} \id_2 + \Pi_{k-j} \big)
\end{equation}
where $\Pi_m$ is 
\begin{equation}
    \Pi_{m} =  \int_{-\pi}^{\pi}\frac{\dd k}{2 \pi}\eE^{-2\ir m k} \ \begin{pmatrix}
-f(k,t)&-g(k,t)\\
 -g(k,t)^*&f(k,t) \\
\end{pmatrix}
\end{equation}
with
\begin{equation}
\begin{split}
f(k,t) &= \sin k \sin(4 \cos(k) t), \\
g(k,t) &= \eE^{-\ir k}(\cos k + \ir \sin k \cos(4 \cos(k) t)).
\end{split}
\end{equation}
\end{subequations}

We stress that the N\'eel and dimer states are not the only states that satisfy these properties, but we focus on them for their simplicity.

\subsection{Charged moments from correlation matrices}

For the two quenches we consider, the RDM of $A$ is a Gaussian operator and can be obtained from the correlations matrix $C_A(t)=[C(t)]_{x,x'}$ with $x,x' \in A$ \cite{cp-01,p-03,pe-09}. We introduce the matrix $J_{A_1\cup A_2}=2 C_A-\mathbb{I}$ and denote its eigenvalues by $\nu_j$. Because of the geometry of $A$, $J_{A_1 \cup A_2}$ has the following block structure, 
\begin{equation}
J_{A_1\cup A_2} = \begin{pmatrix}
J_{11} & J_{12} \\
J_{21} & J_{22}\end{pmatrix} ,
\end{equation}
where $J_{ij}$ is a matrix of size $\ell_i \times \ell_j$ which contains the correlations between sites in $A_i$ and $A_j$. The exact form for the entries are given in Eqs. \eqref{eq:CNeel} and \eqref{eq:CDimer} for the quench from the N\'eel and the dimer state, respectively. From this matrix we introduce $J_{\pm}$ and $J_{\mathrm{x}}$, with respective eigenvalues $\nu^{\pm}_j$ and $\nu^{\mathrm{x}}_j$, as follows, 
\begin{subequations}
\begin{equation}
\label{eq:Jpmdef}
    J_{\pm} =  \begin{pmatrix}
-J_{11} & \pm \ir J_{12} \\
\pm \ir J_{21} & J_{22}\end{pmatrix},
\end{equation}
and
\begin{equation}
\label{eq:Jx}
J_{\mathrm{x}} =(\id + J_+ J_-)^{-1}\cdot(J_++J_-).
\end{equation}
\end{subequations}
Following Refs. \cite{cgs-18,mbc-21,sr-19, paola,gec-18}, we recover the formula for the charged moments $N_{n_e}(\alpha)$ with even $n_e$ defined in Eq. \eqref{eq:supalpha},
\begin{multline}
\label{eq:NnalphaFf}
    \log N_{n_e}(\alpha)=-\ir  \frac{\ell \alpha}{2}+ \sum_{j=1}^{\ell} \log\Big[ \Big(\frac{1+\nu^\mathrm{x}_j}{2}\Big)^{\frac{n_e}{2}}\eE^{\ir \alpha}+\Big(\frac{1-\nu^\mathrm{x}_j}{2}\Big)^{\frac{n_e}{2}}\Big]+\frac{n_e}{2} \sum_{j=1}^{\ell} \log\Big[ \Big(\frac{1+\nu_j}{2}\Big)^{2}+\Big(\frac{1-\nu_j}{2}\Big)^{2}\Big]. 
\end{multline}
The limit for $n_e \rightarrow 1$ gives the charged logarithmic negativity,
\begin{equation}
\label{eq:EalphaFf}
\mathcal{E}(\alpha)=-\ir  \frac{\ell \alpha}{2}+\sum_{j=1}^{\ell} \log\Big[ \Big(\frac{1+\nu^\mathrm{x}_j}{2}\Big)^{1/2}\eE^{\ir \alpha}+\Big(\frac{1-\nu^\mathrm{x}_j}{2}\Big)^{1/2}\Big]+\frac{1}{2} \sum_{j=1}^{\ell} \log\Big[ \Big(\frac{1+\nu_j}{2}\Big)^{2}+\Big(\frac{1-\nu_j}{2}\Big)^{2}\Big].
\end{equation}

For odd $n_o$, we introduce two additional matrices, 
\begin{subequations}
\begin{equation}
J_{n_o} =\Big[ (\id+J_\mathrm{x})^{\frac{n_o-1}2}+ (\id-J_\mathrm{x})^{\frac{n_o-1}2}\Big]^{-1} \cdot \Big[ (\id+J_\mathrm{x})^{\frac{n_o-1}2}- (\id-J_\mathrm{x})^{\frac{n_o-1}2} \Big], \\
\end{equation}
and
\begin{equation}
J_\alpha =\Big[ \id + \frac{\eE^{\ir \alpha}-1}{\eE^{\ir \alpha}+1} J_+ \Big]^{-1} \cdot \Big[J_+ +  \frac{\eE^{\ir \alpha}-1}{\eE^{\ir \alpha}+1} \id \Big], 
\end{equation}
\end{subequations}
and find the expression
\begin{multline}
\log N_{n_o}(\alpha) =-\ir  \frac{\ell \alpha}{2}+ \sum_{j=1}^{\ell}\log \Big[ \frac{1+\nu^+_j}{2} \eE^{\ir \alpha}+\frac{1-\nu^+_j}{2}\Big] + \log\Big[ \det \Big( \frac{\id + J_{n_o} \cdot J_\alpha}{2}\Big)\Big]\\
+\sum_{j=1}^{\ell} \log\Big[ \Big(\frac{1+\nu^\mathrm{x}_j}{2}\Big)^{\frac{n_o-1}2}+\Big(\frac{1-\nu^\mathrm{x}_j}{2}\Big)^{\frac{n_o-1}2}\Big]+\frac{n_o-1}{2} \sum_{j=1}^{\ell} \log\Big[ \Big(\frac{1+\nu_j}{2}\Big)^{2}+\Big(\frac{1-\nu_j}{2}\Big)^{2}\Big]. 
\end{multline}
Even though we used standard techniques from the algebra of Gaussian operators to derive this formula for odd values of $n_o$, it is, to the best of our knowledge, the first time it appears in the literature. The limit $n_o \to 1$ yields the known result from \cite{mbc-21} for the charged probability $N_1(\alpha)$,
\begin{equation}
\label{eq:N1alphaFf}
    \log N_1(\alpha)=-\ir  \frac{\ell \alpha}{2}+\sum_{j=1}^{\ell}\log \Big[ \frac{1+\nu^+_j}{2} \eE^{\ir \alpha}+\frac{1-\nu^+_j}{2}\Big].
\end{equation}

\section{Quench dynamics of the charged R\'enyi logarithmic negativities}\label{sec:Nna}

In this section, we present our analytical results and conjectures for the charged R\'enyi logarithmic negativities $\mathcal{E}_n(\alpha)= \log N_n(\alpha)$ after the quenches from the N\'eel and dimer states. The starting point in the computations is the exact expression for the two-point correlation matrix given in Eqs. \eqref{eq:CNeel} and \eqref{eq:CDimer} for the respective quenches.

\subsection[Analytical results for $\mathcal{E}_1(\alpha)$]{Analytical results for $\boldsymbol{\mathcal{E}_1(\alpha)}$}

To perform analytical calculations, it is useful to express the charged probability $N_1(\alpha)$ in \eqref{eq:N1alphaFf} as a Taylor series in $\Tr J_+^m$, where $J_+$ is defined in Eq. \eqref{eq:Jpmdef}. We introduce the function $h_{n,\alpha}(x)$ and the coefficients $c_{n,\alpha}(m)$ as
\begin{equation}
\label{eq:hna}
h_{n,\alpha}(x) = \log \left[ \left(\frac{1+x}{2} \right)^n \eE^{\ir \alpha}+ \left( \frac{1 - x}{2}\right)^n \right] \equiv \sum_{m=0}^{\infty}c_{n,\alpha}(m) x^m
\end{equation}
and conclude
\begin{equation}
\label{eq:N1_sum}
\log N_1(\alpha) = -\ir  \frac{\ell \alpha}{2}+ \sum_{m=0}^{\infty}c_{1,\alpha}(m) \Tr J_+^m.
\end{equation}

\begin{figure}
\begin{center}
\includegraphics[scale=0.25]{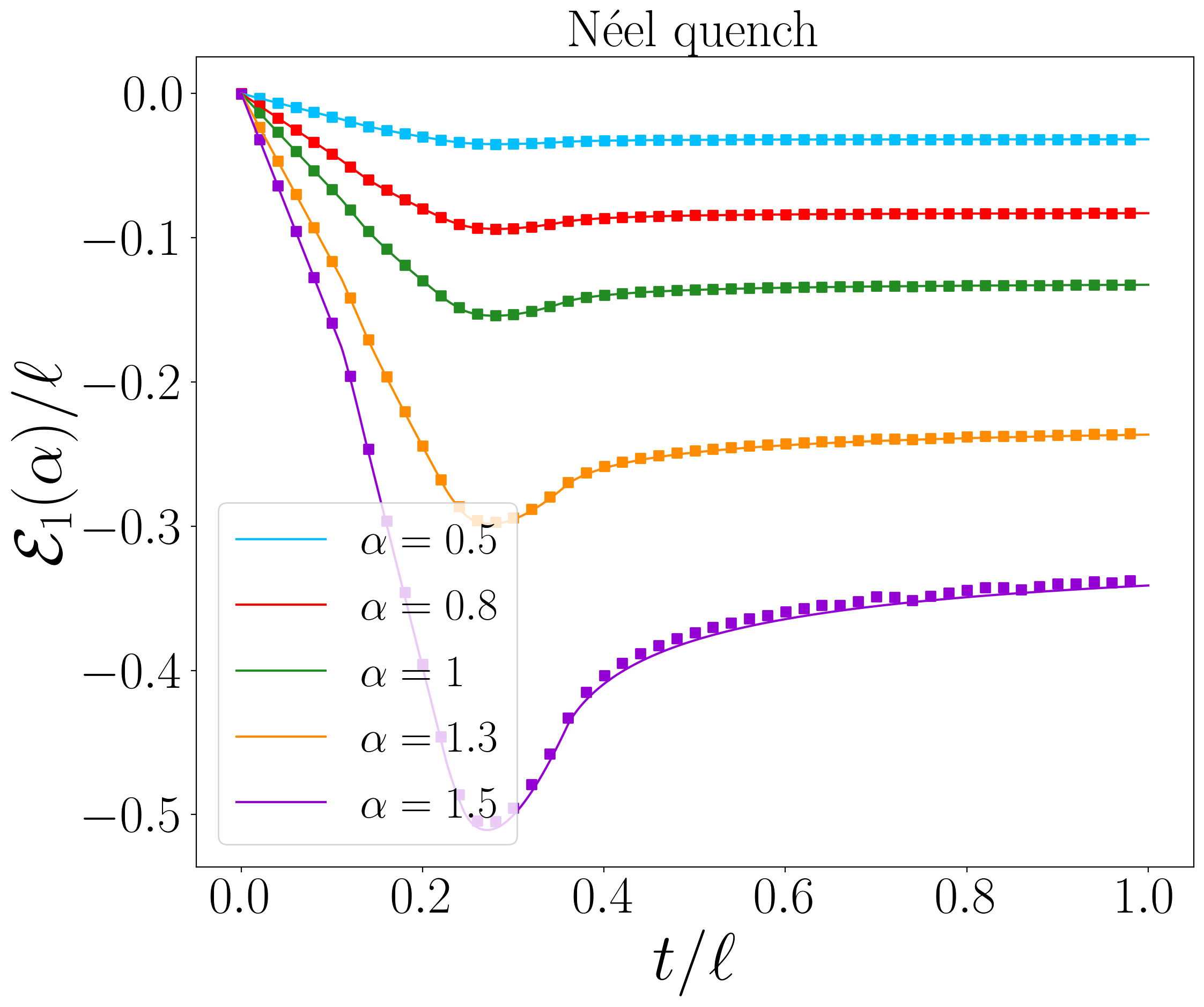}
\includegraphics[scale=0.25]{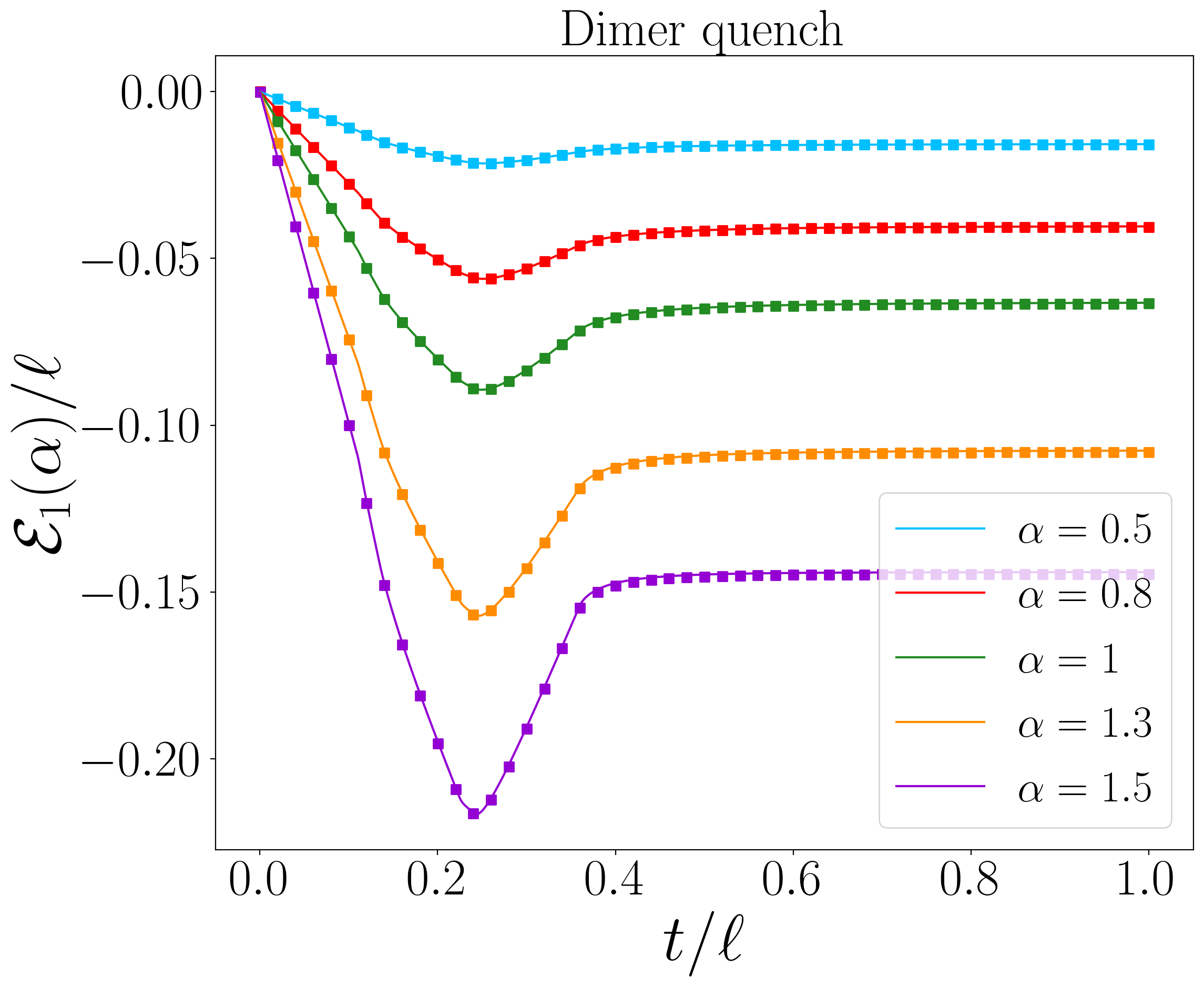}
\end{center}
\caption{Time evolution of $\mathcal{E}_1(\alpha)$ after a quench from the N\'eel state (left) and the dimer state (right) in the tight-biding model \eqref{eq:Hfree} as a function of $t/\ell$ with $\ell_1=230$, $\ell_2=270$ and $d=220$. The analytical prediction of Eq.~\eqref{eq:N1_ex} (solid lines) perfectly matches the numerical data (symbols).}
\label{fig:LogN1}
\end{figure}

We use the stationary phase approximation discussed in \cite{fc-08} and generalise the procedure to the case where the subsystem $A$ consists of two disjoint intervals
 in the {\it scaling limit}  where $\ell_1,\ell_2,d,t \to \infty$ with fixed ratios $\ell_1/\ell$, $d/\ell$ and $t/\ell$. 
 This non-trivial extension of \cite{fc-08} allows us to derive new analytical results in the context of non-equilibrium disjoint systems, and in particular we find \cite{pb-22}
\begin{equation}
\label{eq:TrJ+}
\begin{split}
\Tr J_+(t)^{2j}&= \ell -  \int \frac{\dd k}{2 \pi}(1-x_k^{2j})(\min(\ell_1,2 v_k t)+\min(\ell_2,2 v_k t))\\
&+ \int \frac{\dd k}{2 \pi}\sigma(x_k,j) (\max(d, 2v_kt)+ \max(d+\ell, 2v_kt)- \max(d+\ell_1, 2v_k t)- \max(d+\ell_2, 2v_k t)), \\
\Tr J_+(t)^{2j+1}&=0,
\end{split}
\end{equation}
where  
\begin{equation}
\sigma(x_k,j) =  \sum_{s=1}^j \sum_{t=0}^{2j-2s}(-1)^s x_k^{2j-2 s} (1-x_k^2)^s \frac{ s j}{(2j-s-t)(s+t)}\begin{pmatrix}
2j-s-t \\ s
\end{pmatrix} \begin{pmatrix}
s+t \\ s
\end{pmatrix},
\end{equation}
and $v_k=2 |\sin k|$. The maximal value of the velocity is $v_{\max}=\max_k v_k=2$. The variable $x_k$ depends on the quench, and we have 
\begin{equation}
\label{eq:x_k}
x_k = \begin{cases}
0, & \textrm{N\'eel}, \\
\cos k, & \textrm{dimer}.
\end{cases}
\end{equation}
For the N\'eel case, there is an important simplification, since $\sigma(0,j)= (-1)^j$. The re-summation of~\eqref{eq:TrJ+} into Eq. \eqref{eq:N1_sum} yields
\begin{multline}
\label{eq:N1_ex}
\mathcal{E}_1(\alpha) =\int \frac{\dd k}{2 \pi}\mathrm{Re}[h_{1,\alpha}(x_k)](\min(\ell_1,2 v_k t)+\min(\ell_2,2 v_k t))\\
-\int \frac{\dd k}{2 \pi}\mathrm{Re}[h_{1,\alpha}(x_k)-\frac 12 h_{1,2\alpha}(x_k) ] (\max(d, 2v_kt)+ \max(d+\ell, 2v_kt)- \max(d+\ell_1, 2v_k t)- \max(d+\ell_2, 2v_k t)).
\end{multline}

We compare this analytical prediction with numerical results in Fig. \ref{fig:LogN1} and find a very good agreement for both quenches. 

\subsection[Conjectures for $\mathcal{E}_n(\alpha)$ with arbitrary $n$]{Conjectures for $\boldsymbol{\mathcal{E}_n(\alpha)}$ with arbitrary $\boldsymbol{n}$}

\begin{figure}[t]
\begin{center}
\includegraphics[scale=0.25]{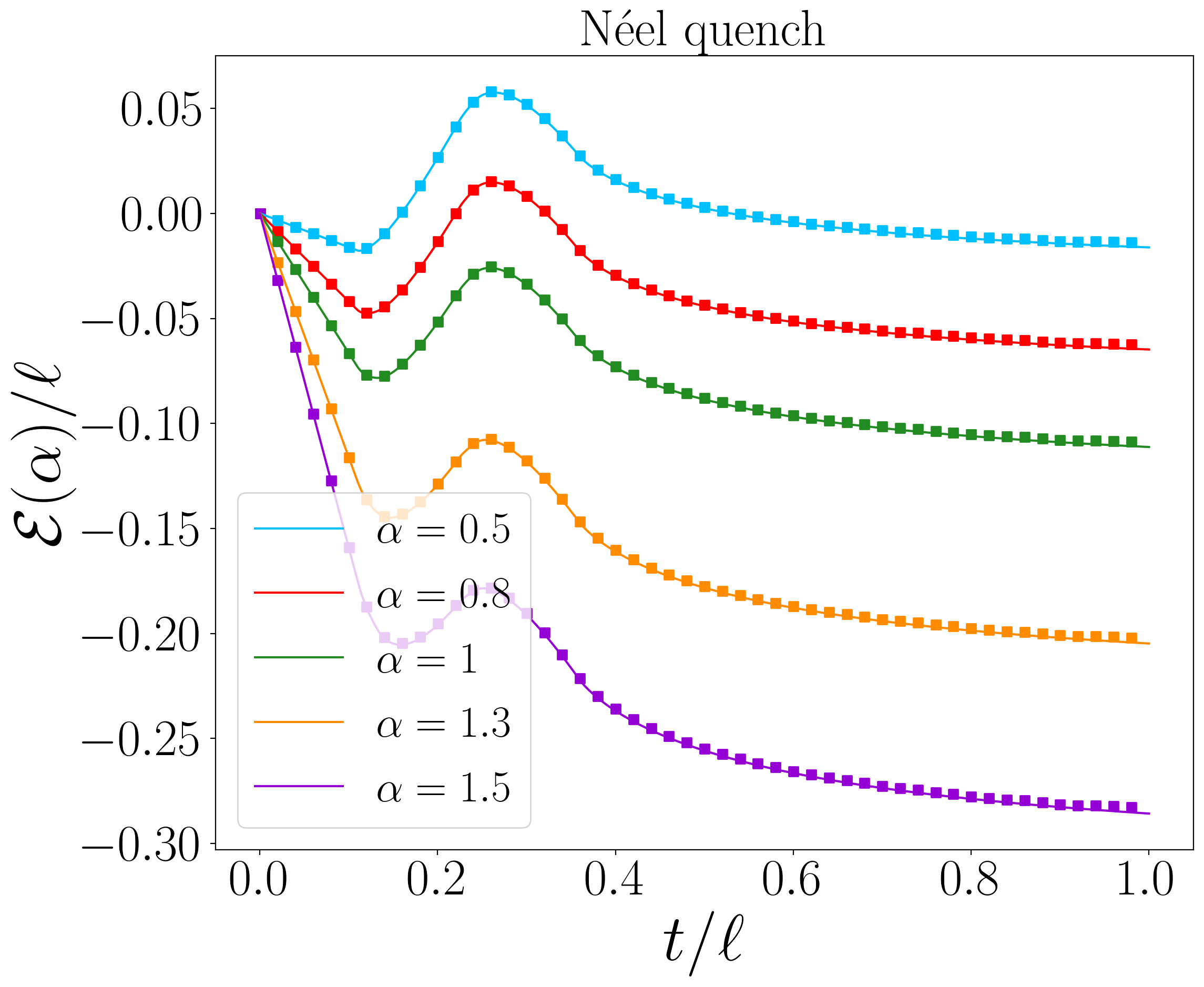}
\includegraphics[scale=0.25]{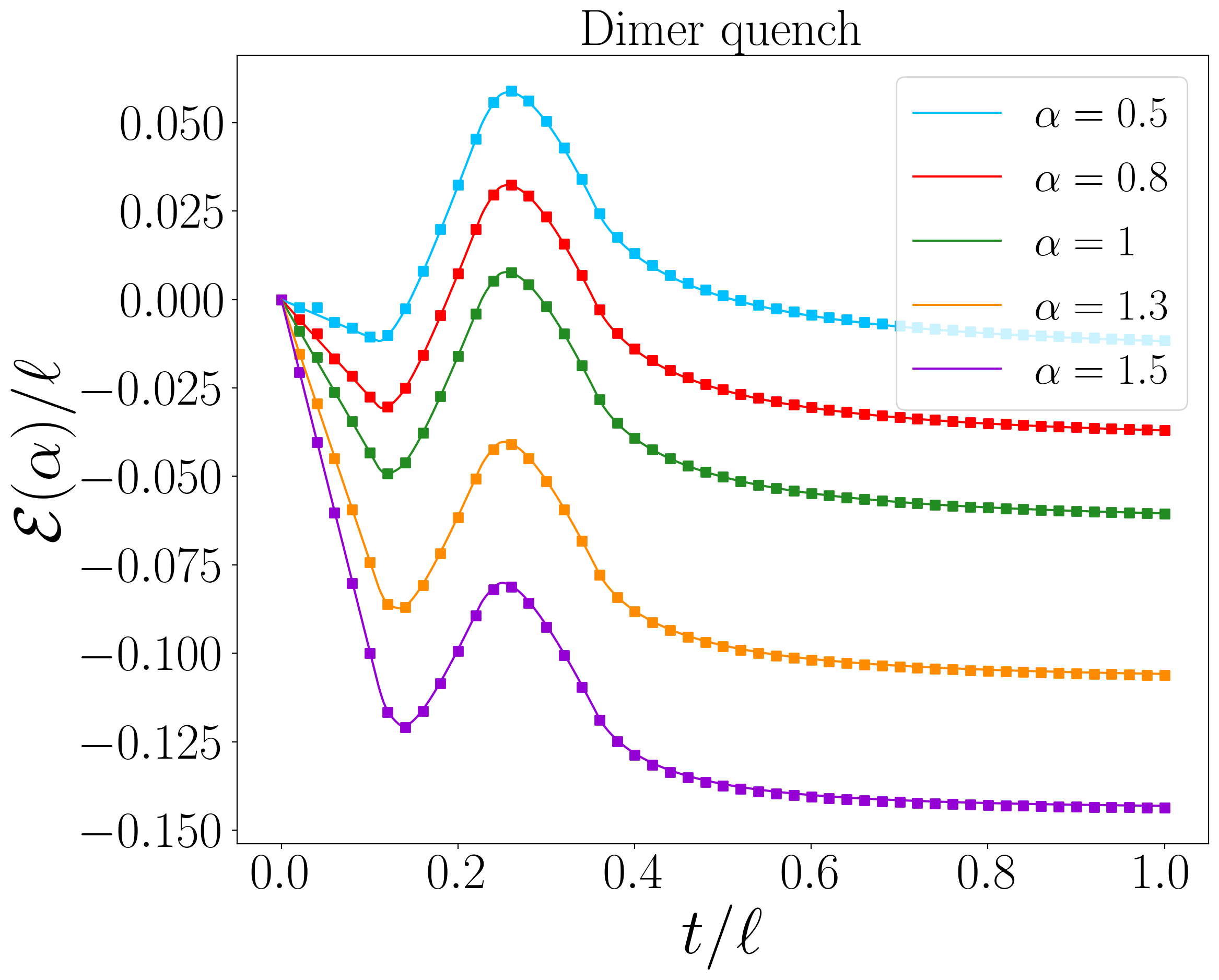}
\end{center}
\caption{Time evolution of $\mathcal{E}(\alpha)$ after a quench from the N\'eel state (left) and the dimer state (right) in the tight-biding model \eqref{eq:Hfree} as a function of $t/\ell$ with $\ell_1=230$, $\ell_2=270$ and $d=220$. The conjecture of Eq.~\eqref{eq:EaConj} (solid lines) perfectly matches the numerical data (symbols).}
\label{fig:E}
\end{figure}

In the cases where $n \neq 1$, we conjecture 
\begin{multline}
\label{eq:LogNConj}
\mathcal{E}_n(\alpha) =\int \frac{\dd k}{2 \pi}\mathrm{Re}[h_{n,\alpha}(x_k)](\min(\ell_1,2 v_k t)+\min(\ell_2,2 v_k t))\\
-\int \frac{\dd k}{2 \pi}\mathrm{Re}[h_{n,\alpha}(x_k)- h^{(2)}_{n,\alpha}(x_k) ] (\max(d, 2v_kt)+ \max(d+\ell, 2v_kt)- \max(d+\ell_1, 2v_k t)- \max(d+\ell_2, 2v_k t))
\end{multline}
with
\begin{equation}
h^{(2)}_{n,\alpha}(x_k)= 
\begin{cases}
\frac{1}{2}h_{n,2 \alpha}(x_k), & \mathrm{odd} \ n, \\[.3cm]
h_{\frac n2, \alpha}(x_k), & \mathrm{even} \ n.  
\end{cases}
\end{equation}

This formula reduces to the exact result \eqref{eq:N1_ex} for $n=1$, and matches quasiparticle conjectures for the R\'enyi logarithmic negativities in the limit $\alpha=0$ \cite{mac-21}. In particular, for the charged logarithmic negativity, the limit $n_e \to 1$ in the even case yields 
\begin{multline}
\label{eq:EaConj}
\mathcal{E}(\alpha) =\int \frac{\dd k}{2 \pi}\mathrm{Re}[h_{1,\alpha}(x_k)](\min(\ell_1,2 v_k t)+\min(\ell_2,2 v_k t))\\
-\int \frac{\dd k}{2 \pi}\mathrm{Re}[h_{1,\alpha}(x_k)-h_{1/2,\alpha}(x_k)] (\max(d, 2v_kt)+ \max(d+\ell, 2v_kt)- \max(d+\ell_1, 2v_k t)- \max(d+\ell_2, 2v_k t)). 
\end{multline}

\begin{figure}[t]
\begin{center}
\includegraphics[scale=0.25]{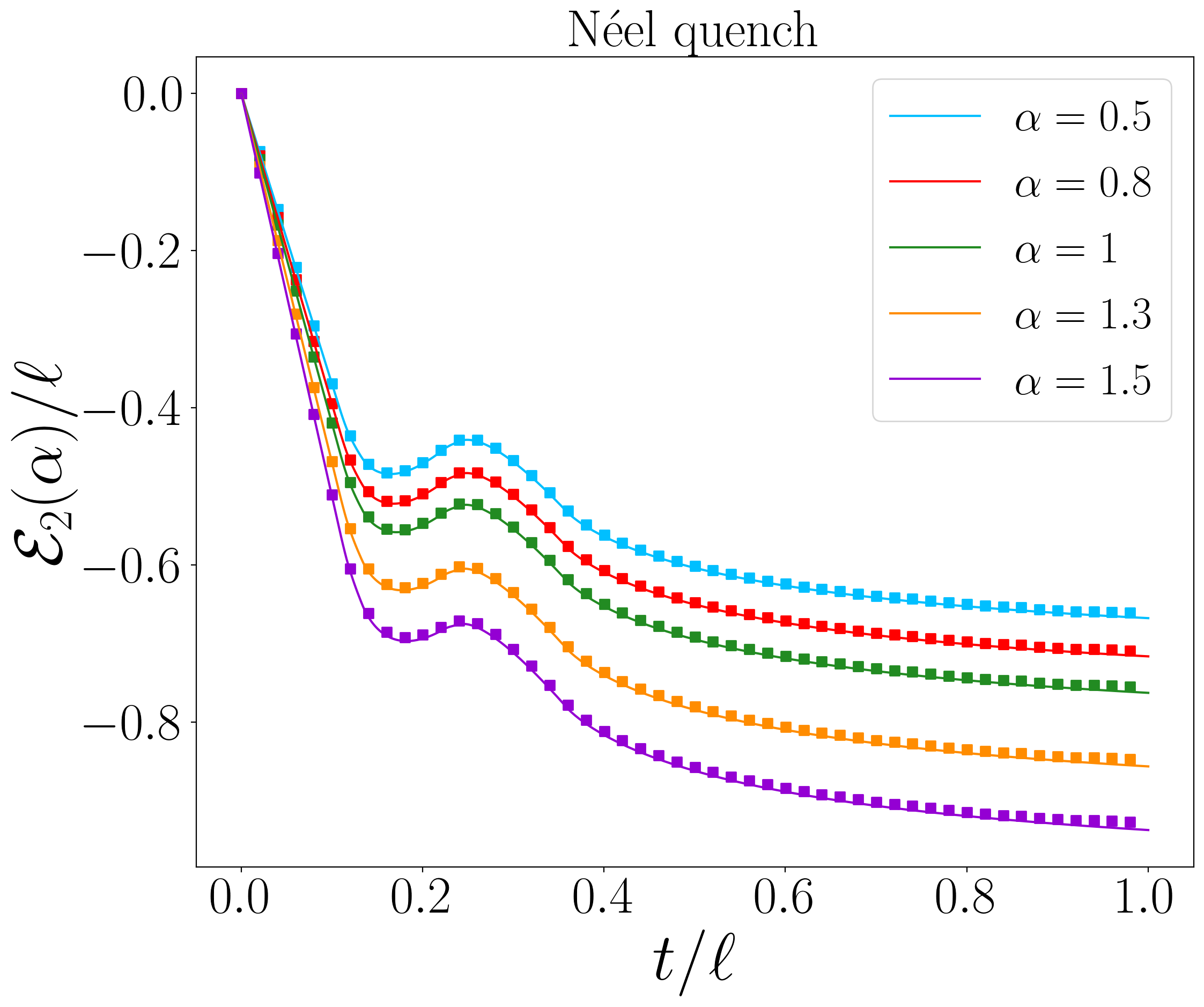}
\includegraphics[scale=0.25]{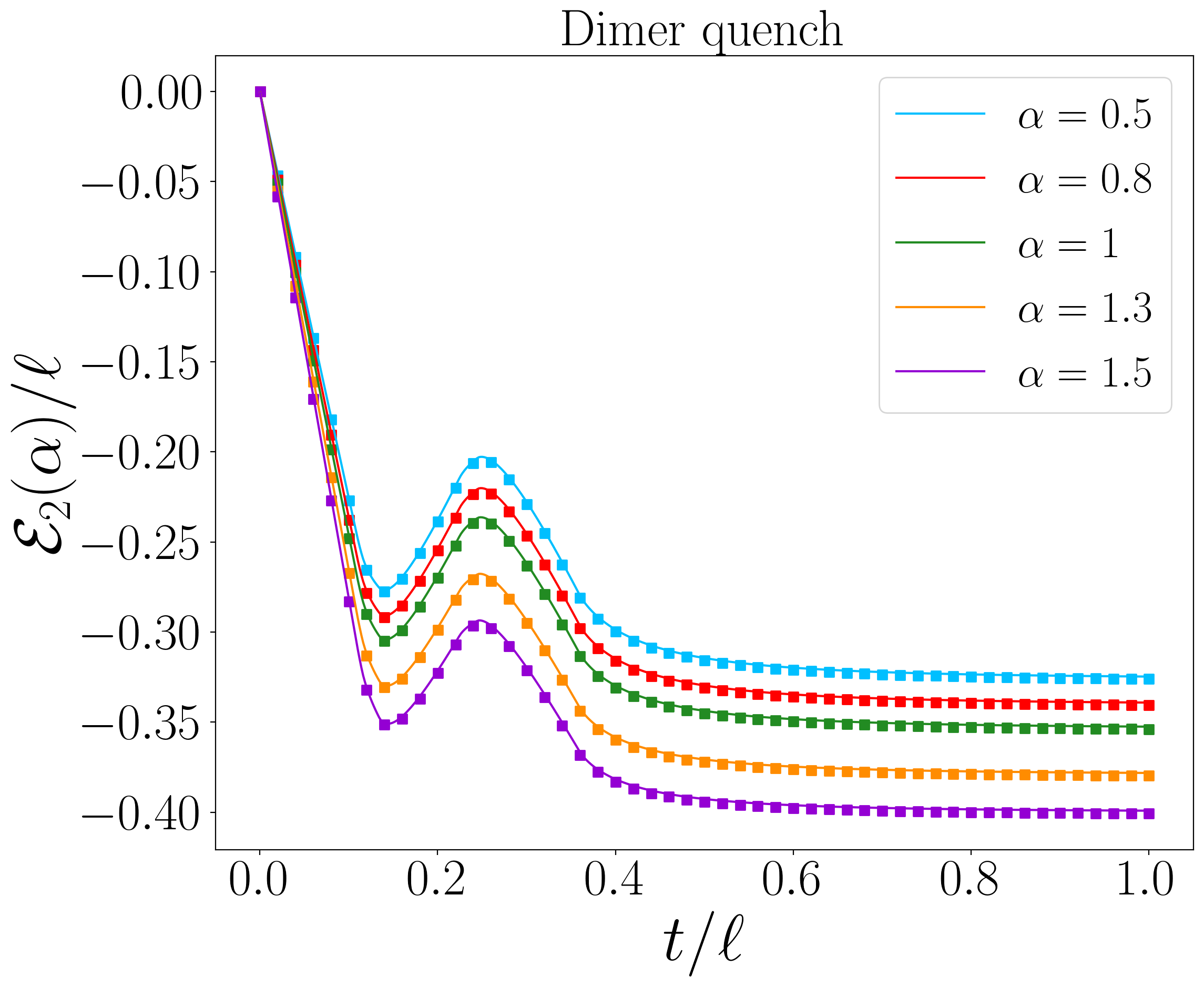}
\includegraphics[scale=0.25]{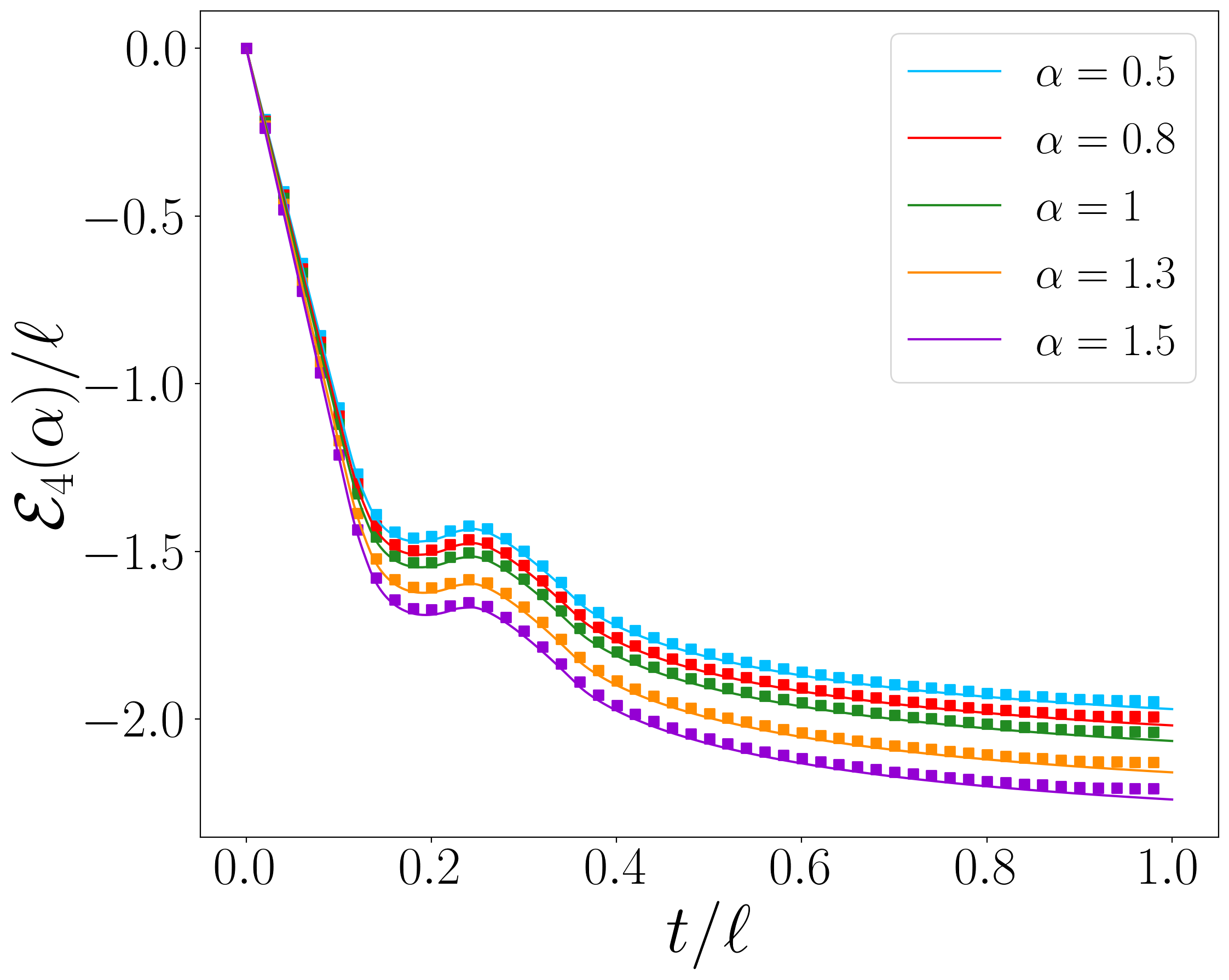}
\includegraphics[scale=0.25]{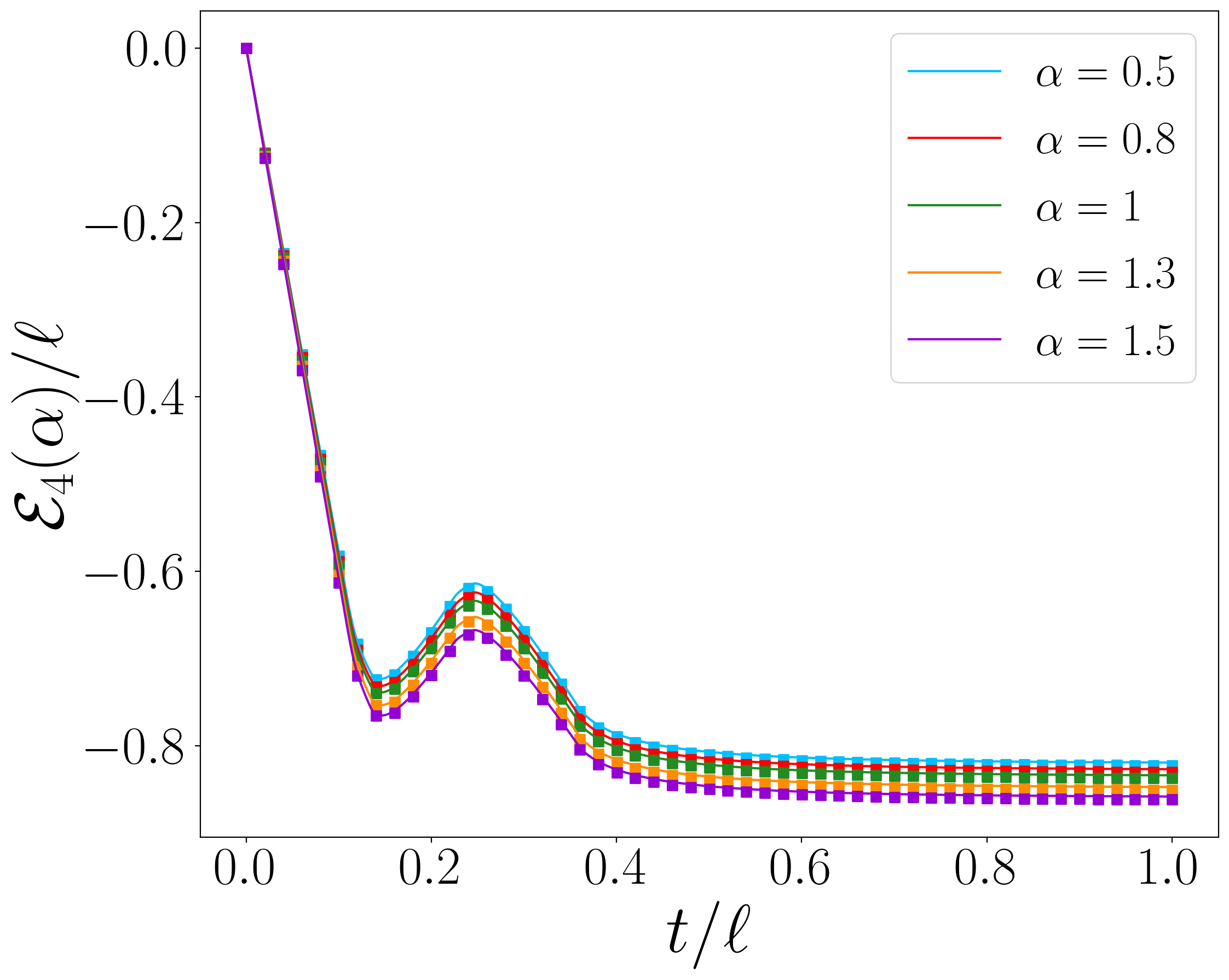}
\end{center}
\caption{Time evolution of $\mathcal{E}_{n_e}(\alpha)$ with $n_e=2,4$ after a quench from the N\'eel state (left) and the dimer state (right) in the tight-biding model \eqref{eq:Hfree} as a function of $t/\ell$ with $\ell_1=230$, $\ell_2=270$ and $d=220$. The conjecture of Eq.~\eqref{eq:LogNConj} (solid lines) perfectly matches the numerical data (symbols).}
\label{fig:LogNe}
\end{figure}

\begin{figure}[t]
\begin{center}
\includegraphics[scale=0.25]{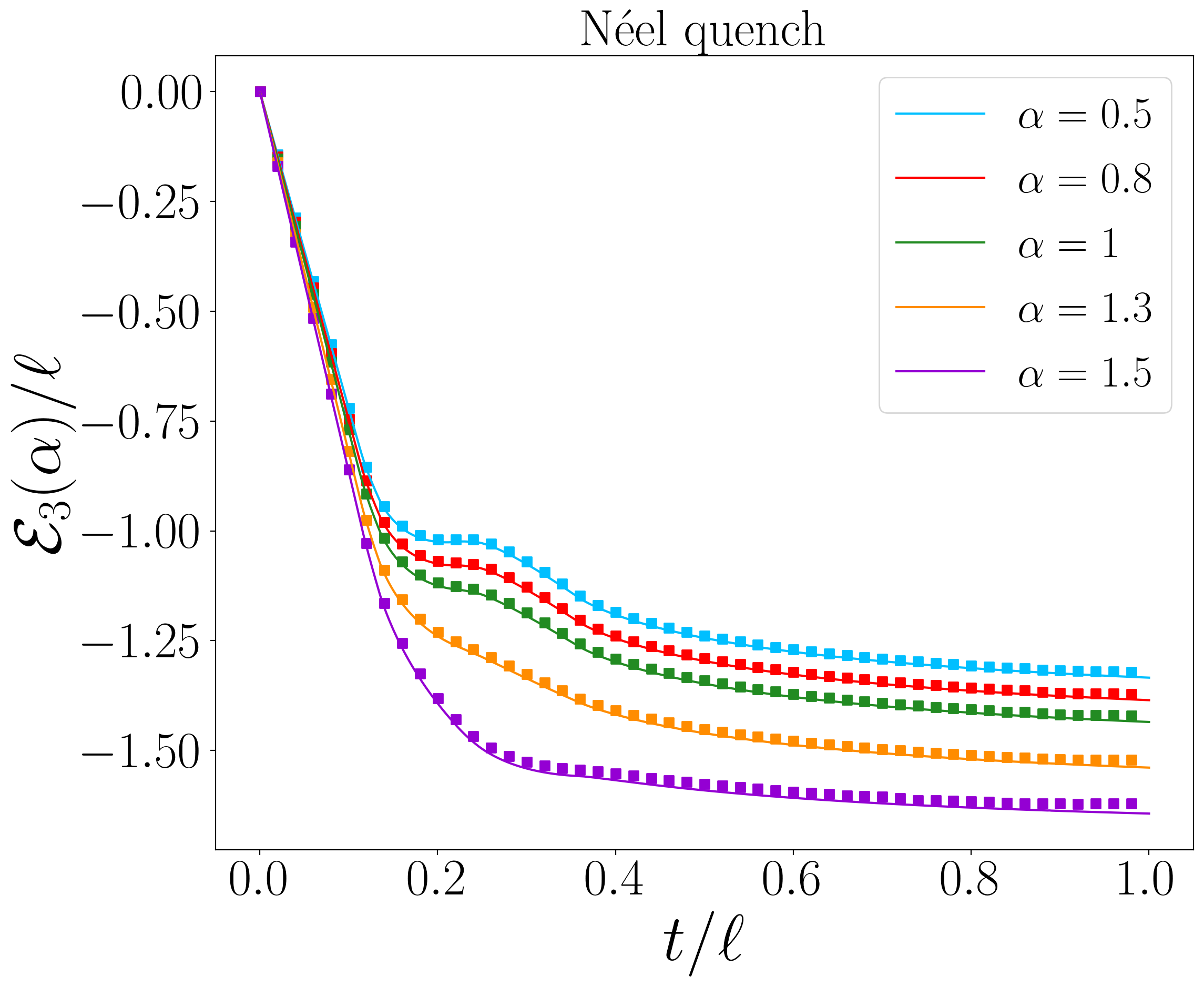}
\includegraphics[scale=0.25]{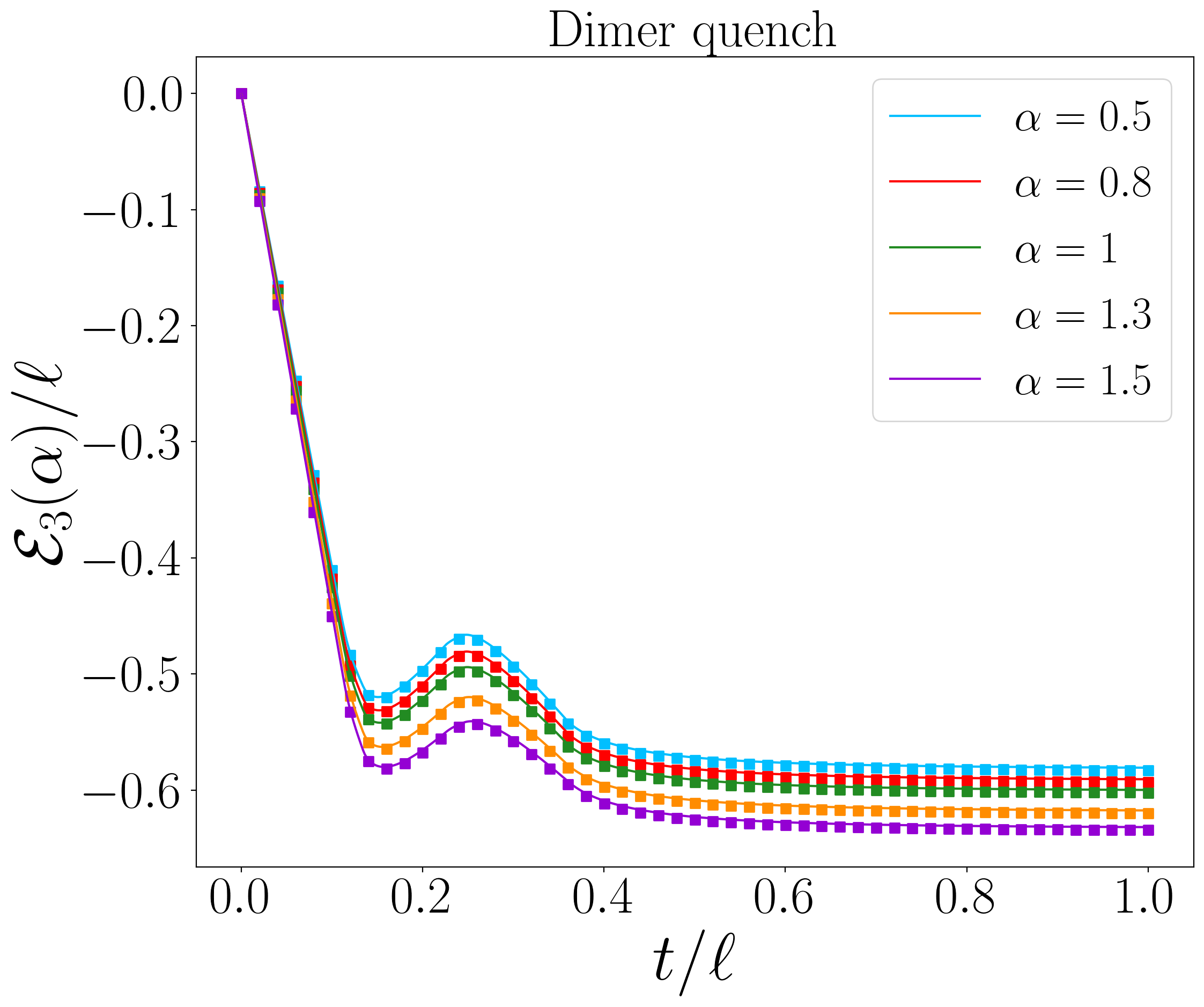}
\includegraphics[scale=0.25]{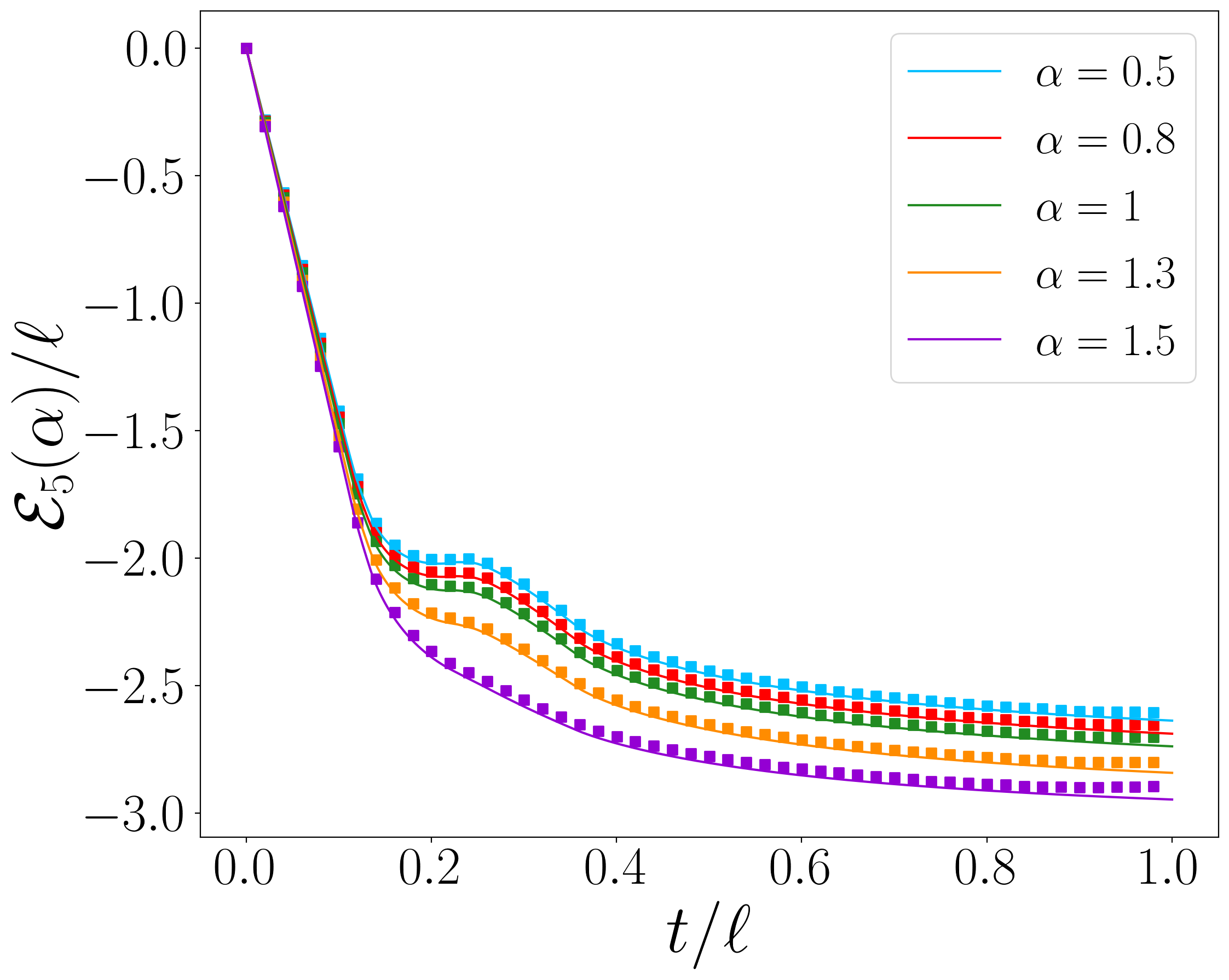}
\includegraphics[scale=0.25]{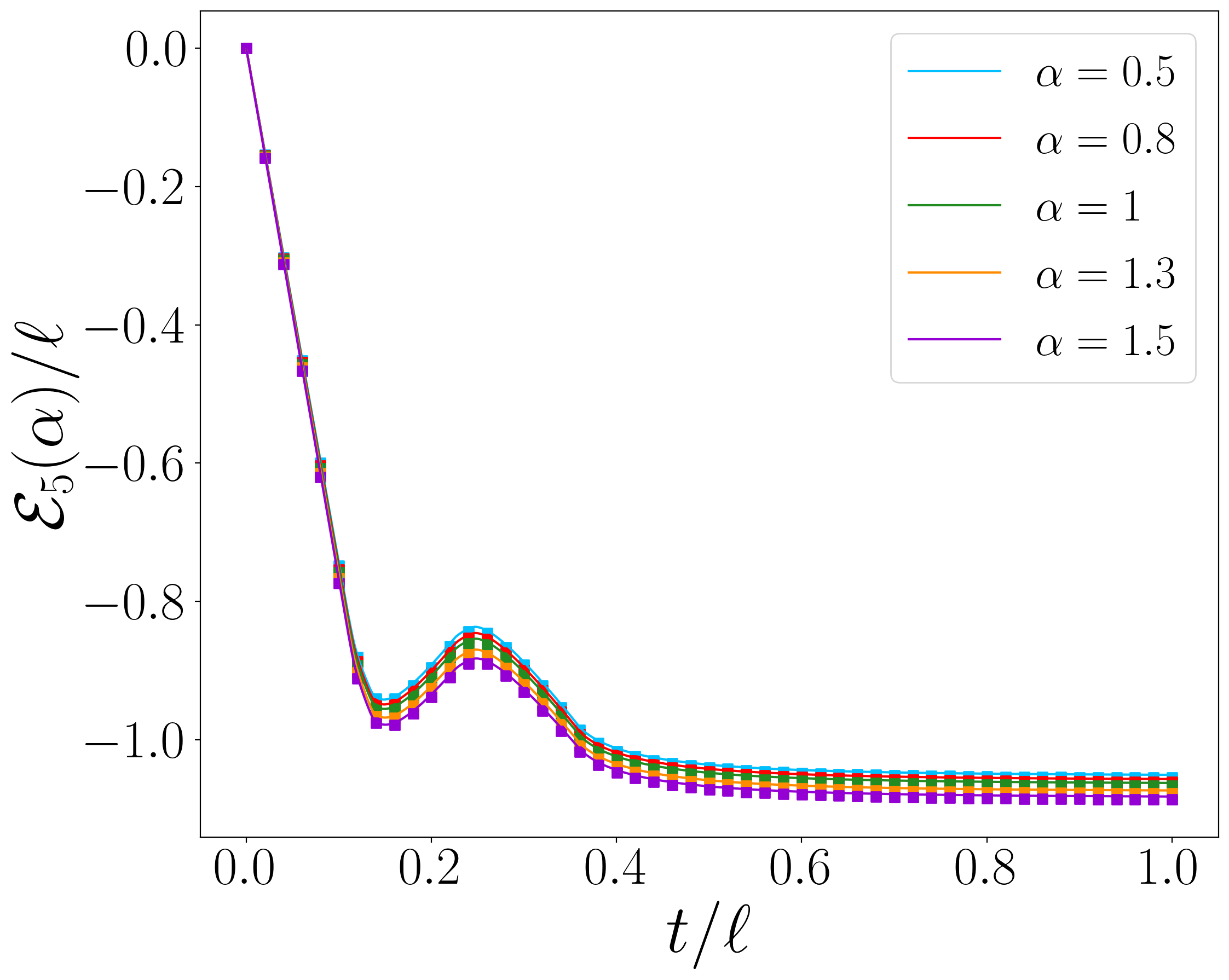}
\end{center}
\caption{Time evolution of $\mathcal{E}_{n_o}(\alpha)$ with $n_o=3,5$ after a quench from the N\'eel state (left) and the dimer state (right) in the tight-biding model \eqref{eq:Hfree} as a function of $t/\ell$ with $\ell_1=230$, $\ell_2=270$ and $d=220$. The conjecture of Eq.~\eqref{eq:LogNConj} (solid lines) perfectly matches the numerical data (symbols).}
\label{fig:LogNo}
\end{figure}

We compare the conjectures for the charged logarithmic negativity and $\mathcal{E}_n(\alpha)$ for even and odd $n$ in Figs. \ref{fig:E}, \ref{fig:LogNe} and \ref{fig:LogNo}, respectively. We systematically find a very good agreement between the conjectures \eqref{eq:LogNConj}, \eqref{eq:EaConj}, and the numerical results, for both quenches.

\section{Charge-imbalance-resolved negativity}\label{sec:cir_t}

In this section, we investigate the charge-imbalance-resolved negativity after the two quenches under consideration. Looking at \eqref{eq:NegDefNeel}, we need to compute the Fourier transform of the charged logarithmic negativity $\mathcal{E}(\alpha)$ and the charged probability $N_1(\alpha)$ to obtain $\mathcal{Z}_{R_1}(q)$ and $p(q)$, given in Eqs. \eqref{eq:ZR1} and \eqref{eq:pqft}, respectively. We also recover known results for the quench dynamics of the total logarithmic negativity from the charge-imbalance-resolved ones. 

\subsection{Fourier transforms}

To compute the Fourier transforms, we approximate the charged negativity and probability at quadratic order in $\alpha$. We introduce the integrals
\begin{equation}
\label{eq:J}
\begin{split}
\J_{A_1,A_2}^{(1)} &= \int \frac{\dd k}{2 \pi}(1-x_k^2)(\min(\ell_1,2 v_k t)+\min(\ell_2,2 v_k t)), \\
\J_m^{(1)} &= \int \frac{\dd k}{2 \pi}(1-x_k^2) (\max(d, 2v_kt)+ \max(d+\ell, 2v_kt)- \max(d+\ell_1, 2v_k t)- \max(d+\ell_2, 2v_k t)), \\
\J_m^{(1/2)} &= \int \frac{\dd k}{2 \pi}\frac{2 }{1+(1-x_k^2)^{-1/2}} (\max(d, 2v_kt)+ \max(d+\ell, 2v_kt)- \max(d+\ell_1, 2v_k t)- \max(d+\ell_2, 2v_k t)),
\end{split}
\end{equation}
and find 
\begin{equation}
\begin{split}
\mathcal{E}(\alpha) &=  \mathcal{E}(0)-\frac{\alpha^2}{8}\big(\J_{A_1,A_2}^{(1)}-\J_m^{(1)}+\J_m^{(1/2)} \big) + \mathcal{O}(\alpha^4),\\[.5cm]
\mathcal{E}_1(\alpha) &= -\frac{\alpha^2}{8}\big(\J_{A_1,A_2}^{(1)}+\J_m^{(1)} \big) + \mathcal{O}(\alpha^4),
\end{split}
\end{equation}
where
\begin{equation}
\label{eq:E0}
\mathcal{E}(0)=\int \frac{\dd k}{2 \pi}h_{1/2,0}(x_k)(\max(d, 2v_kt)+ \max(d+\ell, 2v_kt)- \max(d+\ell_1, 2v_k t)- \max(d+\ell_2, 2v_k t)).
\end{equation}
From the definitions \eqref{eq:def_log_neg} and \eqref{eq:def_Ea}, this quantity is the total logarithmic negativity, $\mathcal{E}(0)\equiv \mathcal{E}$. 

Next, we evaluate the Fourier transform of these quantities with respect to the charge-imbalance operator $\hat{Q}_A=Q_1+Q_2-\ell/2$ with charge $q=\Delta q + \langle \hat{Q}_A \rangle$. In both quenches, we have $\langle \hat{Q}_A \rangle = 0$, and the integrals yield
\begin{align}
\label{eq:Zq}
\mathcal{Z}_{R_1}(q) &=\eE^{\mathcal{E}(0) } \int_{-\pi}^{\pi}\frac{d \alpha}{2 \pi} \eE^{-\ir  \Delta q \alpha} \eE^{-\frac{\alpha^2}{8}(\J_{A_1,A_2}^{(1)}-\J_m^{(1)}+\J_m^{(1/2)})}\\ \nonumber
&=\eE^{\mathcal{E}(0) } \eE^{-\frac{2 \Delta q^2}{\J_{A_1,A_2}^{(1)}-\J_m^{(1)}+\J_m^{(1/2)}}}\sqrt{\frac{2}{\pi(\J_{A_1,A_2}^{(1)}-\J_m^{(1)}+\J_m^{(1/2)})}} , \\[.3cm]
p(q) &=  \int_{-\pi}^{\pi}\frac{d \alpha}{2 \pi} \eE^{-\ir  \Delta q \alpha} \eE^{-\frac{\alpha^2}{8}(\J_{A_1,A_2}^{(1)}+\J_m^{(1)})} \label{eq:pq} \\ \nonumber
&= \eE^{-\frac{2 \Delta q^2}{\J_{A_1,A_2}^{(1)}+\J_m^{(1)}}}\sqrt{\frac{2}{\pi(\J_{A_1,A_2}^{(1)}+\J_m^{(1)})}}.
\end{align}

We note that for the N\'eel quench  there is a closed-form expression for $\mathcal{Z}_{R_1}(q)$ in terms of Gamma functions, similar to the charged entropies given in \cite{pbc-21bis} for the same quench. We do not report this expression here, since it yields the same results as Eq. \eqref{eq:Zq} in the limit $|\Delta q| \ll \ell$. Moreover, there is no closed-form expression for $p(q)$, and hence we de not have an exact result for the charge-imbalance-resolved logarithmic negativity $\hat{\mathcal{E}}(q)$.

\subsection{Charge-imbalance-resolved logarithmic negativity}

\begin{figure}
\begin{center}
\includegraphics[scale=0.25]{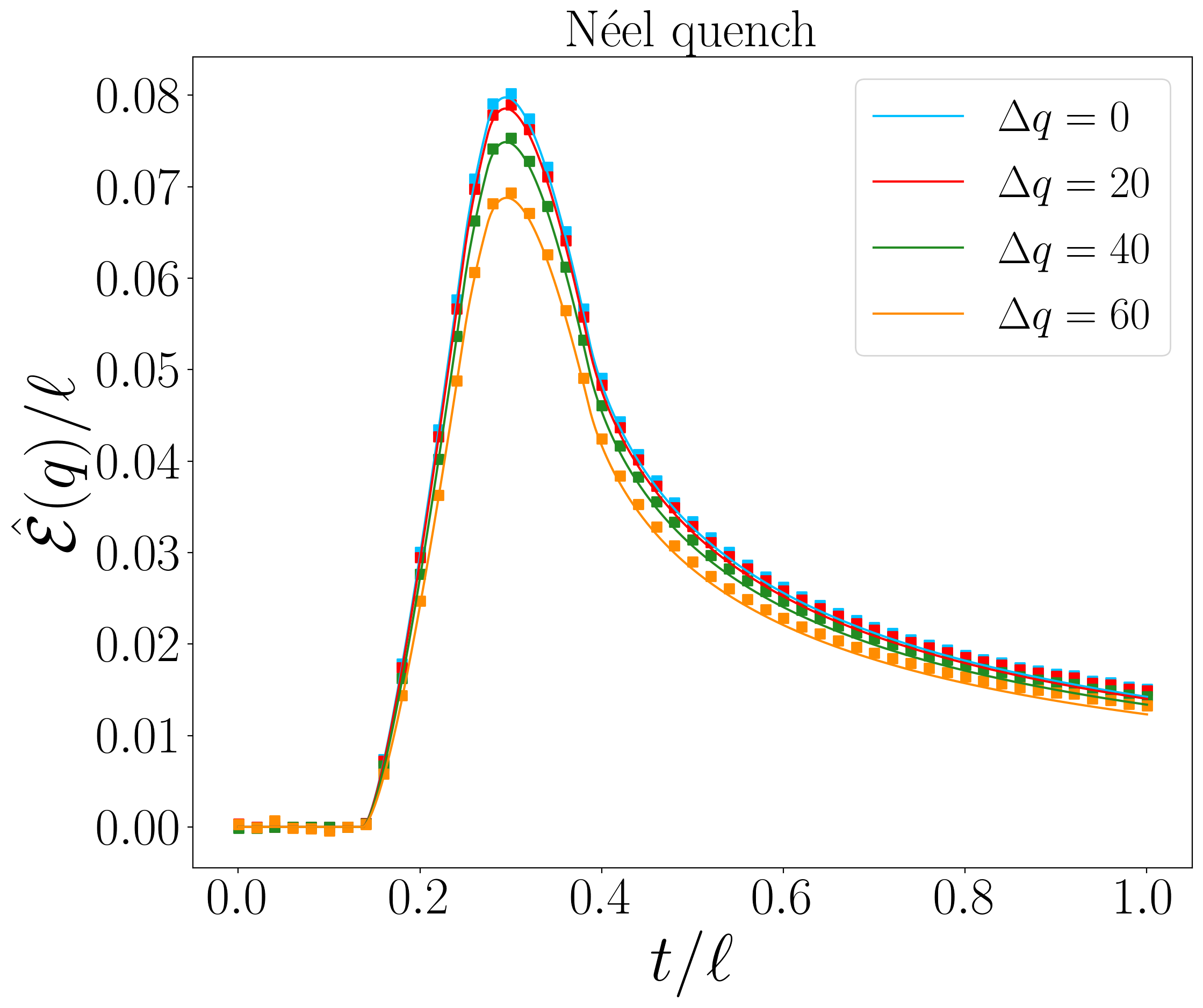}
\includegraphics[scale=0.25]{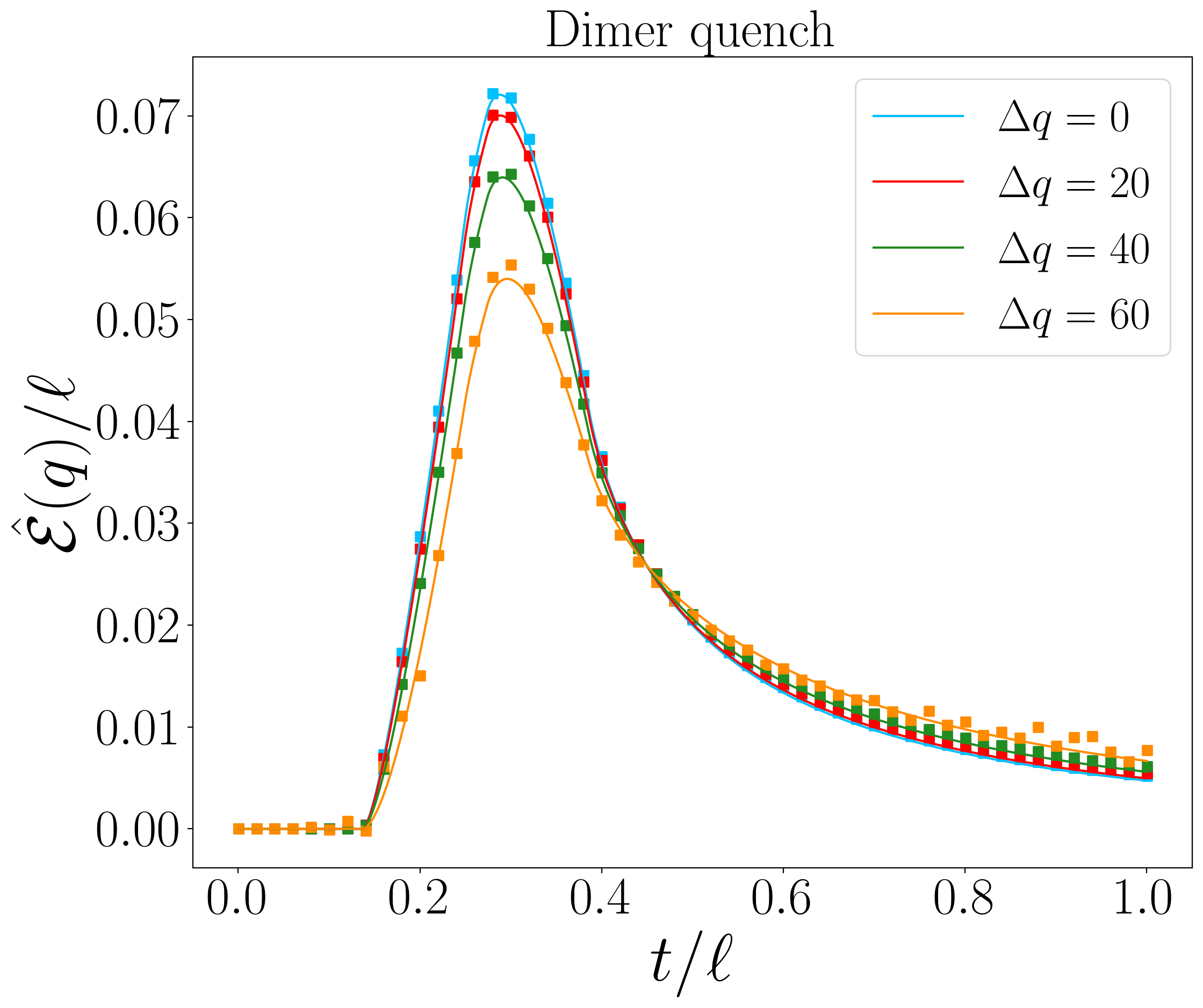}
\end{center}
\caption{Time evolution of $\hat{\mathcal{E}}(q)$ after a quench from the N\'eel state (left) and the dimer state (right) in the tight-biding model \eqref{eq:Hfree} as a function of $t/\ell$ with $\ell_1=180$, $\ell_2=220$ and $d=220$. The approximation of Eq.~\eqref{eq:Eq_approx} (solid lines) matches the numerical data (symbols) convincingly for small values of $|\Delta q|$ and start deviating from the data as $|\Delta q|$ increases considerably.}
\label{fig:Eq}
\end{figure}

We insert Eqs. \eqref{eq:Zq} and \eqref{eq:pq} in \eqref{eq:NegDefNeel} and find
\begin{equation}
\label{eq:Eq_approx}
\hat{\mathcal{E}}(q) = \mathcal{E}(0) -2 \Delta q^2 \Big( \frac{1}{\J_{A_1,A_2}^{(1)}-\J_m^{(1)}+\J_m^{(1/2)}}-\frac{1}{\J_{A_1,A_2}^{(1)}+\J_m^{(1)}} \Big) + \frac 12 \log \Big( \frac{\J_{A_1,A_2}^{(1)}+\J_m^{(1)}}{\J_{A_1,A_2}^{(1)}-\J_m^{(1)}+\J_m^{(1/2)}}\Big).
\end{equation}
We stress that this result holds in the limit $|\Delta q| \ll \ell$. We compare the prediction of Eq. \eqref{eq:Eq_approx} with ab initio numerical results in Fig. \ref{fig:Eq} and find a convincing match for small values of $|\Delta q|$. To understand the qualitative behaviour of the charge-imbalance-resolved negativity, let us discuss the $\J$-integrals in Eq. \eqref{eq:J}. We suppose without loss of generality that $\ell_1 \leqslant \ell_2$. The integral $\J_{A_1,A_2}^{(1)}$ has two distinct regimes. For $t \leqslant\ell_1/(2 v_{\max})$, it grows linearly with $t$, whereas it saturates to a value proportional to $\ell$ in the limit $t \to \infty$. The integrals $\J_m^{(1)}$ and $\J_m^{(1/2)}$ both vanish for $t \leqslant d/(2 v_{\max})$ and in the limit $t \to \infty$, and reach a value proportional to $\ell$ for intermediate times. Looking at \eqref{eq:Eq_approx}, we thus conclude that there is an equipartition broken at order $\Delta q^2/\ell$ for intermediate times, whereas the equipartition is exact for $t \leqslant d/(2 v_{max})$ and in the limit $t \to \infty$. These behaviours are well reproduced by our numerical analysis in Fig. \ref{fig:Eq}, and are reminiscent of the quench dynamics of the symmetry-resolved mutual information discussed in \cite{pbc-21bis}. This is not surprising, since it is known that the total logarithmic negativity and mutual information have a similar behaviour out of equilibrium \cite{ac-19bis, BKL22}. 

\subsection{Total negativity}

To check the consistency of our results, we wish to recover the total negativity from the charge-imbalance-resolved ones. We recast Eq. \eqref{eq:Negtot} as
\begin{equation}
\label{eq:Negtot2}
\mathcal{N}= \sum_{q=-\ell/2}^{\ell/2} p(q)\frac{\eE^{\hat{\mathcal{E}}(q)}-1}{2}.
\end{equation}
For clarity, we consider the two sums in the right-hand side separately. With the quadratic approximations of Eqs. \eqref{eq:pq} and \eqref{eq:Eq_approx}, we have 
\begin{equation}
\label{eq:pqSums}
\begin{split}
  \sum_{q=-\ell/2}^{\ell/2} p(q) &=\sqrt{\frac{2}{\pi(\J_{A_1,A_2}^{(1)}+\J_m^{(1)})}} \sum_{\Delta q=-\ell/2}^{\ell/2} \eE^{-\frac{2 \Delta q^2}{\J_{A_1,A_2}^{(1)}+\J_m^{(1)}}}, \\[.3cm]
  \sum_{q=-\ell/2}^{\ell/2} p(q)\eE^{\hat{\mathcal{E}}(q)} &= \eE^{\mathcal{E}(0) }  \sqrt{\frac{2}{\pi(\J_{A_1,A_2}^{(1)}-\J_m^{(1)}+\J_m^{(1/2)})}} \sum_{\Delta q=-\ell/2}^{\ell/2} \eE^{-\frac{2 \Delta q^2}{\J_{A_1,A_2}^{(1)}-\J_m^{(1)}+\J_m^{(1/2)}}}.
\end{split}
\end{equation}
We note that the quadratic approximations are valid for $|\Delta q| \ll \ell$, and this condition is not met for the extreme values of $\Delta q$ in the sum. However, these values for the subsystem charge have a very low probability, and we nonetheless expect these approximations to give good results. Similar approximations provided excellent results for the total entanglement entropy, see \cite{pbc-21bis}. Both terms in Eq. \eqref{eq:pqSums} have the form $\sqrt{\frac{2}{\pi\J}} \sum_{\Delta q=-\ell/2}^{\ell/2} \eE^{-\frac{2 \Delta q^2}{\J}}$ for some $\J>0$. In the large-$\ell$ limit, we approximate the sums as integrals over the real axis, and use the Gaussian integral
\begin{equation}
 \sqrt{\frac{2}{\pi\J}} \int_{-\infty}^{\infty}\dd \Delta q \ \eE^{-\frac{2 \Delta q^2}{\J}} =1 .
\end{equation}
This simplifies both terms in Eq. \eqref{eq:pqSums}. With Eq. \eqref{eq:Negtot2}, we conclude 
\begin{equation}
\mathcal{N} = \frac{\eE^{\mathcal{E}(0)}-1}{2},
\end{equation}
where $\mathcal{E}(0)$ is the total logarithmic negativity given in Eq. \eqref{eq:E0}. We thus recover the known quench dynamics for the total logarithmic negativity from the charge-imbalance-resolved ones \cite{ac-19bis}. 

\section{Quasiparticle picture for the charged R\'enyi logarithmic negativities}\label{sec:QPP}

In this section, we discuss a physical picture, known as the \textit{quasiparticle picture} \cite{cc-05,ac-17,ac-18}, that allows one to predict the time evolution of entanglement measures after a global quantum quench from a low-entangled initial state in a one-dimensional quantum integrable model. In particular, we argue that our results for the charged R\'enyi logarithmic negativities can be understood in terms of the quasiparticle picture, and generalise recent results for the dynamics of R\'enyi logarithmic negativities \cite{mac-21}. 

\subsection{Quasiparticle picture and entanglement dynamics for free fermions}

\begin{figure}
\begin{center}
  \begin{tikzpicture}[>=stealth, scale=0.91]
  
    \foreach \x in {0,2,...,10}
      {
         \fill[orange!40] (\x,0)--(\x+0.9,0.9) arc (45:135:1.275) -- cycle;
         
       \draw[->](\x,0)--(\x-0.5,0.5);
           \draw(\x,0)--(\x-1.15,1.15);
           
           \draw[->](\x,0)--(\x+0.5,0.5);
           \draw(\x,0)--(\x+1.15,1.15);}
           
     \foreach \x in {0,2,...,2}
      \draw[draw=red,fill=red!40] (\x,0) circle (2pt);
      \draw [-,red,very thick] (-0.75,0)--(2.5,0);
      \draw[red,thick] (-1, 0) node{$\cdots$};
      
     \foreach \x in {4,6,...,6}
      \draw[draw=blue,fill=blue!40] (\x,0) circle (2pt);
      \draw [-,blue,very thick] (2.5,0)--(7.5,0);
      
    \foreach \x in {8,10,...,10}
      \draw[draw=red,fill=red!40] (\x,0) circle (2pt);
      \draw [-,red,very thick] (7.5,0)--(10.75,0);
      \draw[red,thick] (11.1, 0) node{$\cdots$};

           \draw[thick](2.5,-0.1)--(2.5,0.1);
           \draw[thick](7.5,-0.1)--(7.5,0.1);
           
           \draw[->] (-1.425,-0.2)--(-1.425,2);
              \draw (-1.225,2) node{$\boldsymbol{t}$};
           
           \draw[blue, thick] (5, -0.3) node{ $\boldsymbol{A}$};

           \draw[<->] (2.5,-1)--(7.5,-1);
            \draw (5, -1.2) node{$\ell$};
        
   \end{tikzpicture}  

\end{center}
\caption{Illustration of the quasiparticle picture for the entanglement spreading in a one-dimensional integrable system. The arrows indicate the propagation of the quasiparticles with the largest velocity~$v_{\max}$.}
\label{fig:QuasiPFi}
\end{figure}
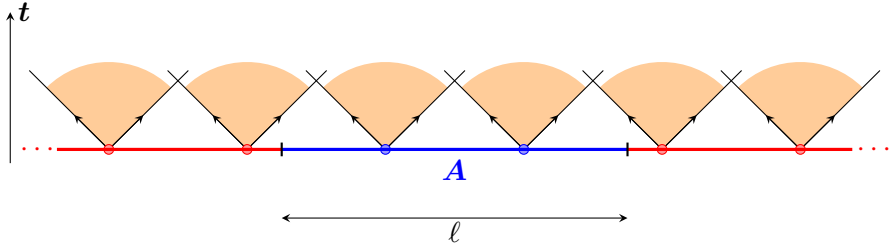

In the global quench protocol, the system is prepared in an initial state $|\psi_0\rangle$ that has an extensive amount of energy compared to the groundstate of the Hamiltonian $H$ which governs the subsequent time evolution. Because of this surplus of energy, the initial state acts as a source of quasiparticle excitations, which are assumed to be produced in independent pairs with opposite momenta $k$ and $-k$. This assumption can be weakened in free models \cite{btc-18,bc-18,bc-20b}, but it is fundamental for interacting integrable systems, as argued in \cite{PPV-17}. After the quench, each particle moves ballistically through the system, with velocity $v_k$. For spin chains with local interactions, the Lieb-Robinson bound \cite{lr-72} guarantees that there is a maximum value $v_{\max}$ for the possible velocities of the quasiparticles. The key element of this description is that the quasiparticles emitted from different points are incoherent, while those emitted from the same point are entangled. The entanglement between a subsystem $A$ and its complement $B$ is thus proportional to the number of entangled pairs of quasiparticles shared between $A$ and $B$. In the case where $A$ is a single interval of length $\ell$ embedded in an infinite line, the quasiparticle picture predicts that the entanglement entropy evolves as \cite{cc-05}
\begin{equation}
    \label{EntropyQuasi}
    S_1(t)= 2 t \int_{2 v_kt<\ell}\dd k \ v_k s(k)+ \ell \int_{2 v_kt>\ell} \dd k \ s(k)
\end{equation}
in the scaling limit $\ell,t \to \infty$ with fixed ratio $t/\ell$. Here, the function $s(k)$ depends on the rate of production of the quasiparticles with momentum $\pm k$ and their contribution to the entanglement entropy. This situation is illustrated in Fig. \ref{fig:QuasiPFi}. The two terms of Eq.~\eqref{EntropyQuasi} give two different regimes as the entropy evolves in time. For times $t\leqslant \ell/(2v_{\max})$, the domain of integration of the second integral vanishes, and the entropy grows linearly in time. For larger times, the entanglement entropy growth slows down, and for $t\gg \ell/(2 v_{\max})$ the entanglement entropy saturates to a value that is extensive in the subsystem size.

To give predictive power to the quasiparticle result \eqref{EntropyQuasi}, one needs to compute the function $s(k)$, which depends strongly on the model and on the initial state. For free-fermion models, this function reads \cite{ac-17}
\begin{equation}
\label{sk}
 s(k) =\frac{1}{2 \pi} (-n_k \log n_k -(1-n_k)\log(1-n_k))
\end{equation}
where $n_k$ is the probability of occupation of the mode $k$ in the stationary state. The result \eqref{EntropyQuasi} also holds for the R\'enyi entropies $S_n$ with the replacement \cite{ac-17b}
\begin{equation}
\label{eq:snk}
   s(k)\to s_n(k)=\frac{1}{2 \pi}\frac{1}{1-n} \log(n_k^n+(1-n_k)^n).
\end{equation}
In terms of the function $h_{n,\alpha}(x)$ defined in Eq. \eqref{eq:hna}, we thus have 
\begin{equation}
    \label{EvEntFree}
S_n(t) =\frac{1}{1-n}\int \frac{\dd k}{2 \pi} h_{n,0}(2n_k-1)\min(\ell,2v_k t)
\end{equation}
where we use the function $\min(\ell,2 v_kt)$ to recast the sum of two integrals in \eqref{EntropyQuasi} in a simpler form. 
%
%
We mention that finding a generalisation of Eq. \eqref{EntropyQuasi} for R\'enyi entropies with $n\neq1$ was a long-standing open problem for interacting integrable models \cite{ac-17b,ac-17c,mac-18,clsv-19,kb-21} which has been solved only very recently \cite{BKALC22}.

\subsection{Quasiparticle dynamics for the charged R\'enyi logarithmic negativities}

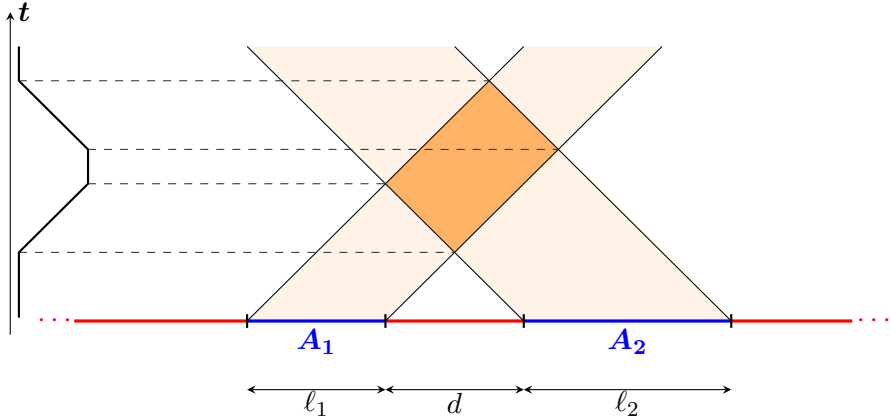
\begin{figure}
\begin{center}
  \begin{tikzpicture}[>=stealth, scale=0.91]

               \fill[orange!10] (2,0)--(6,4)--(8,4)--(4,0)-- cycle;
                \fill[orange!10] (6,0)--(2,4)--(5,4)--(9,0)-- cycle;
                \fill[orange!60] (5,1)--(6.5,2.5)--(5.5,3.5)--(4,2)-- cycle;

      \draw [-,red,very thick] (-0.5,0)--(2,0);
      \draw[red,thick] (-0.75, 0) node{$\cdots$};    
   \draw [-,blue,very thick] (2,0)--(4,0);   
     \draw [-,red,very thick] (4,0)--(6,0);   
       \draw [-,blue,very thick] (6,0)--(9,0);
      \draw [-,red,very thick] (9,0)--(10.75,0);
      \draw[red,thick] (11.1, 0) node{$\cdots$};

           \draw[thick](2,-0.1)--(2,0.1);
           \draw[thick](4,-0.1)--(4,0.1);
            \draw[thick](6,-0.1)--(6,0.1);
             \draw[thick](9,-0.1)--(9,0.1);
           
            \draw[blue, thick] (3, -0.3) node{ $\boldsymbol{A_1}$};
              \draw[blue, thick] (7.5, -0.3) node{ $\boldsymbol{A_2}$};
    
           \draw[<->] (2,-1)--(4,-1);
            \draw (3, -1.2) node{$\ell_1$};
            
             \draw[<->] (4,-1)--(6,-1);
            \draw (5, -1.2) node{$d$};
            
             \draw[<->] (6,-1)--(9,-1);
            \draw (7.5, -1.2) node{$\ell_2$};
            
            \draw (2,0)--(6,4);
            \draw (4,0)--(8,4);
            \draw (6,0)--(2,4);
            \draw (9,0)--(5,4);
            
             \draw[->] (-1.425,-0.2)--(-1.425,4.5);
              \draw (-1.225,4.5) node{$\boldsymbol{t}$};
       
       \draw[thick] (-1.3,0.05)--(-1.3,1)--(-0.3,2)--(-0.3,2.5)--(-1.3,3.5)--(-1.3,4)    ;
       
       \draw[dashed,black!80] (-1.3,1)--(5,1);
         \draw[dashed,black!80] (-0.3,2)--(4,2);
          \draw[dashed,black!80] (-0.3,2.5)--(6.5,2.5);
           \draw[dashed,black!80] (-1.3,3.5)--(5.5,3.5);

   \end{tikzpicture}  
\end{center}
\caption{Illustration of the entanglement dynamics between disjoint intervals in the case where there is a single velocity for the quasiparticles. At a given time $t$, the number of quasiparticles shared between $A_1$ and $A_2$ is proportional to the width of the darker orange region. The typical behaviour of the logarithmic negativity is reported on the vertical plot on the left. It illustrates the delay before the linear increase, the plateau and the linear decrease discussed in the main text.}
\label{fig:QuasiPFiDisjoint}
\end{figure}

Let us consider the case in which the subsystem $A$ of length $\ell$ consists of two disjoint intervals $A_1$ and $A_2$ of respective lengths $\ell_1$ and $\ell_2$ that are separated by a distance $d$. Without loss of generality we consider $\ell_1 \leqslant \ell_2$. In this case, the dynamics of the logarithmic negativity can be understood in terms of the quasiparticle picture. The entanglement between $A_1$ and $A_2$ is proportional to the number of pairs of entangled quasiparticles they share at a given time $t$. To ease the counting exercise, let us first assume that all the quasiparticles have the same velocity $v_k=v$. It follows that for short times $t\leqslant d/(2 v)$, there are no entangled quasiparticles shared by the two intervals and the logarithmic negativity is zero. It then increases linearly for $d/(2 v) \leqslant t \leqslant (d+\ell_1)/(2 v)$. There is a plateau for $(d+\ell_1)/(2 v) \leqslant t \leqslant (d+\ell_2)/(2 v)$, after which it decreases linearly in time up to $t=(d+\ell)/(2v)$. There are no longer any quasiparticles shared between the subsystems for larger times, so that the logarithmic negativity vanishes for $t\geqslant(d+\ell)/(2v)$. We illustrate this behaviour in Fig. \ref{fig:QuasiPFiDisjoint}. Accordingly, the resulting dynamics of the logarithmic negativity is proportional to $(\max(d,2vt)+ \max(d+\ell,2vt)- \max(d+\ell_1,2vt)-\max(d+\ell_2,2vt))$\cite{ctc-14,ac-19}.

In the presence of different velocities $v_k$, the quasiparticle result is \cite{ac-17, ac-18}
\begin{equation}
    \label{eq:logNegQPP}
    \mathcal{E}= \int \frac{\dd k}{2\pi} \ \epsilon(k) (\max(d,2v_kt)+ \max(d+\ell,2v_kt)- \max(d+\ell_1,2v_kt)-\max(d+\ell_2,2v_kt)),
\end{equation}
with $\epsilon(k) =h_{1/2,0}(2n_k-1) $ for free fermions. We note that this is exactly what we find in Eq.~\eqref{eq:E0} for the total logarithmic negativity, with $n_k=1/2$ for the quench from the N\'eel state, and $n_k=(1+\cos k)/2$ for the dimer state.

The quasiparticle picture also describes the dynamics of the R\'enyi logarithmic negativities $\mathcal{E}_n$ defined in \eqref{eq:RN}. The result is \cite{mac-21}
\begin{multline}
\label{eq:logNeqnt}
\mathcal{E}_n =\int \frac{\dd k}{2 \pi}\epsilon_n(k)(\min(\ell_1,2 v_k t)+\min(\ell_2,2 v_k t))\\
-\int \frac{\dd k}{2 \pi}(\epsilon_{n}(k)- \epsilon^{(2)}_{n}(k)) (\max(d, 2v_kt)+ \max(d+\ell, 2v_kt)- \max(d+\ell_1, 2v_k t)- \max(d+\ell_2, 2v_k t))
\end{multline}
with
\begin{equation}
\epsilon^{(2)}_{n}(k)= 
\begin{cases}
\frac{1}{2}\epsilon_{n}(k), & \mathrm{odd} \ n, \\[.3cm]
\epsilon_{\frac n2}(k), & \mathrm{even} \ n.  
\end{cases}
\end{equation}

For free systems, the kernels are 
\begin{equation}
\epsilon_n(k) = h_{n,0}(2n_k-1) =  \log(n_k^n+(1-n_k)^n).
\end{equation}
We note that the limit $n_e \to 1$ of Eq. \eqref{eq:logNeqnt} for even $n_e$ yields Eq. \eqref{eq:logNegQPP}, as expected, because $h_{1,0}(x)=0$.  

Our results of Eqs. \eqref{eq:N1_ex} and \eqref{eq:LogNConj} for the charged R\'enyi logarithmic negativities suggest the quasiparticle conjecture
\begin{multline}
\label{eq:logNeqnt_alpha}
\mathcal{E}_n(\alpha) =\ir \alpha \langle \hat{Q}_A \rangle +\int \frac{\dd k}{2 \pi}\epsilon_{n,\alpha}(k)(\min(\ell_1,2 v_k t)+\min(\ell_2,2 v_k t))\\
-\int \frac{\dd k}{2 \pi}(\epsilon_{n,\alpha}(k)- \epsilon^{(2)}_{n,\alpha}(k)) (\max(d, 2v_kt)+ \max(d+\ell, 2v_kt)- \max(d+\ell_1, 2v_k t)- \max(d+\ell_2, 2v_k t))
\end{multline}
with
\begin{equation}
\label{eq:h2a}
\epsilon^{(2)}_{n,\alpha}(k)= 
\begin{cases}
\frac{1}{2}\epsilon_{n,2\alpha}(k), & \mathrm{odd} \ n, \\[.3cm]
\epsilon_{\frac n2,\alpha}(k), & \mathrm{even} \ n.
\end{cases}
\end{equation}
Here, $\epsilon_{n,0}(k) \equiv \epsilon_n(k)$ from Eq. \eqref{eq:logNeqnt}, and $\epsilon_{n,\alpha}(k) =\mathrm{Re}[h_{n,\alpha}(2n_k-1)]$ for free systems. Even though this conjecture is similar to Eq. \eqref{eq:logNeqnt} and the results of \cite{mac-21}, the $\alpha$-dependence in Eq. \eqref{eq:h2a} is non-trivial and could not have been directly guessed from the results for the total R\'enyi logarithmic negativities. 

We expect that qualitatively similar results apply also to generic integral models. However, the known problems for the calculation of the R\'enyi entropies 
\cite{ac-17b,ac-17c,mac-18} still prevent from the determination of the exact kernels in the integral \eqref{eq:logNeqnt_alpha} that would replace $\epsilon_{n,\alpha}(k)$
and $\epsilon^{(2)}_{n,\alpha}(k)$. The first steps towards the solution of these issues were recently discussed in the literature \cite{PVCC22,BKALC22}.

\section{Conclusion}\label{sec:ccl}

In this paper, we investigated the dynamics of the charge-imbalance-resolved negativity after a quench in a free-fermion chain. We first considered the corresponding charged moments $N_n(\alpha)$, and expressed these quantities in terms of the two-point correlation matrix. Second, we used these formulas to give analytical results and conjectures for the charged R\'enyi logarithmic negativities $\mathcal{E}_n(\alpha)$. We tested those results against ab initio numerical computations for two distinct quenches, and found a systematic very good agreement. Third, we studied the Fourier transforms of the charged moments approximated at quadratic order in $\alpha$ to investigate the charge-imbalance-resolved negativity. Our results show a perfect equipartition for early and large times that is broken at order $\Delta q^2/\ell$ for intermediate ones. These results hold in the limit $|\Delta q| \ll \ell$, and match numerical results with a satisfactory precision. Finally, we argued that our results for the charged R\'enyi logarithmic negativities can be understood in the framework of the quasiparticle picture for the entanglement dynamics, and we provided a conjecture for $\mathcal{E}_n(\alpha)$ that we expect to hold for a large variety of integrable models 
with a proper adaptation. 

There are several avenues that would be worth investigating in the future. First, it would be natural to test our quasiparticle conjecture for the charged R\'enyi logarithmic negativities $\mathcal{E}_n(\alpha)$ (as well as the conjectures for the charged entropies and mutual information, see \cite{pbc-21,pbc-21bis}) in the context of interacting integrable models. A second idea is to investigate symmetry-resolved entanglement measures after an inhomogeneous quench, similarly to the total entanglement entropy in Refs. \cite{alba-inh,BFPC18,ABF19,a-19,dsvc-17,rcdd-19}. Finally, it would be necessary to understand whether one could compute symmetry-resolved entanglement measure for non-integrable models with some of the methods developed in e.g. Refs. \cite{NRVH:17,nvh-18,zn-20,BKP:entropy,GoLa19,bc-20,PBCP20,mac-20,cdc-18,fcdc-19,rpv-19,CZLT20}.

\section*{Acknowledgements}

We acknowledge support from ERC under Consolidator grant number 771536 (NEMO). GP holds a CRM-ISM postdoctoral fellowship and acknowledges support from the Mathematical Physics Laboratory of the CRM. He also thanks SISSA for hospitality during the early stages of this project. RB~acknowledges support from the Croatian Science Foundation (HrZZ) project No. IP-2019-4-3321.

\end{document}